\documentclass[12pt]{article}
\usepackage{frankstyle}

\begin{document}

\title{A Generalized Focused Information Criterion for GMM\footnote{We thank Manuel Arellano, Otilia Boldea, Camilo Garc\'{i}a-Jimeno, Bruce Hansen, Frank Kleibergen, and seminar participants at the 2013 Latin American Workshop in Econometrics, the 2014 Midwest Econometrics Group meetings, Tilburg, the Tinbergen Institute, and the University of Wisconsin for helpful comments.}} 

\author{Minsu Chang \hspace{1em} Francis J.\ DiTraglia}

\date{\footnotesize Final Version: September 16, 2017}

\maketitle 
\begin{abstract}
  \singlespacing
This paper proposes a criterion for simultaneous GMM model and moment selection: the generalized focused information criterion (GFIC). 
Rather than attempting to identify the ``true'' specification, the GFIC chooses from a set of potentially mis-specified moment conditions and parameter restrictions to minimize the mean-squared error (MSE) of a user-specified target parameter.
The intent of the GFIC is to formalize a situation common in applied practice.
An applied researcher begins with a set of fairly weak ``baseline'' assumptions, assumed to be correct, and must decide whether to impose any of a number of stronger, more controversial ``suspect'' assumptions that yield parameter restrictions, additional moment conditions, or both.
Provided that the baseline assumptions identify the model, we show how to construct an asymptotically unbiased estimator of the asymptotic MSE to select over these suspect assumptions: the GFIC. 
We go on to provide results for post-selection inference and model averaging that can be applied both to the GFIC and various alternative selection criteria.
To illustrate how our criterion can be used in practice, we specialize the GFIC to the problem of selecting over exogeneity assumptions and lag lengths in a dynamic panel model, and show that it performs well in simulations.
We conclude by applying the GFIC to a dynamic panel data model for the price elasticity of cigarette demand.


\end{abstract}

\section{Introduction}

An econometric model is a tool for answering a particular research question: different questions may suggest different models for the same data. 
And the fact that a model is wrong, as the old saying goes, does not prevent it from being useful. 
This paper proposes a novel selection criterion for GMM estimation that takes both of these points to heart: the generalized focused information criterion (GFIC). 
Rather than attempting to identify the correct specification, the GFIC chooses from a set of potentially mis-specified moment conditions and parameter restrictions to yield the smallest mean squared error (MSE) estimator of a user-specified scalar target parameter. 
We derive the GFIC under local mis-specification, using asymptotic mean squared error (AMSE) to approximate finite-sample MSE. 
In this framework mis-specification, while present for any fixed sample size, disappears in the limit so that asymptotic variance and squared bias remain comparable. 
GMM estimators remain consistent under local mis-specification but their limit distributions show an asymptotic bias. 
Adding an additional moment condition or imposing a parameter restriction generally reduces asymptotic variance but, if incorrectly specified, introduces a source of bias.
The GFIC trades off these two effects in the first-order asymptotic expansion of an estimator to approximate its finite sample behavior.

The GFIC takes its motivation from a situation that is common in empirical practice.
A researcher who hopes to estimate a parameter of interest $\mu$ must decide which assumptions to use.
On the one hand is a set of relatively uncontroversial ``baseline'' assumptions.
We suppose that the baseline assumptions are correct and identify $\mu$.
But the very fact that the baseline assumptions do not raise eyebrows suggests that they may not be especially informative about $\mu$. 
On the other hand are one or more stronger controversial ``suspect'' assumptions.
These stronger assumptions are expected to be much more informative about $\mu$.
If we were certain that they were correct, we would definitely choose to impose them in estimation.
Indeed, by continuity, even if they were \emph{nearly} correct, imposing the suspect assumptions could yield a favorable bias-variance tradeoff.
This is the essential idea behind the GFIC.
When the baseline assumptions identify the model, the GFIC provides an asymptotically unbiased estimator of AMSE.

The GFIC is an extension of the focused moment selection criterion (FMSC) of \cite{DiTraglia2016}.
While the FMSC considers the problem of selecting moment conditions while holding the model specification \emph{fixed}, the GFIC allows us to select over both aspects of our specification simultaneously.
This extension is particularly valuable in panel data applications, where we may, for example, wish to carry out selection over the lag specification as well as the exogeneity assumptions used to estimate a dynamic panel model.
We specialize the GFIC to such a dynamic panel example below, and provide simulation evidence of its performance.
Online Appendix \ref{sec:additional} provides two additional examples: selecting between random and fixed effects estimators, and choosing between pooled OLS and mean-group estimators of an average effect in the presence of heterogeneity.  
In addition to extending the FMSC to a broader class of problems, we also extend the results of \cite{DiTraglia2016} on post-selection and moment-averaging estimators to the more general setting of the GFIC.  
We conclude with an empirical example modelling the price elasticity of cigarette demand. 

As its name suggests, the GFIC is related to the focused information criterion (FIC) of \cite{ClaeskensHjort2003}, a model selection procedure for maximum likelihood estimators that uses local mis-specification to approximate the MSE of a target parameter. 
Like the FIC and related proposals, e.g.\ \cite{Schorfheide2005}, the GFIC uses local mis-specification to derive a risk-based selection criterion.
Unlike them, however, the GFIC provides both moment and model selection for general GMM estimators. 
If the moment conditions used in estimation are the score of a maximum likelihood model and we consider model selection only, then the GFIC reduces to the FIC.
Thus, the GFIC extends both the FIC and the FMSC of \cite{DiTraglia2016}.
Comparatively few papers propose criteria for simultaneous GMM model and moment selection under mis-specification.\footnote{See \cite{Smith1992} for an approach to GMM model selection based on non-nested hypothesis testing. For a detailed discussion of the literature on moment selection, see \cite{DiTraglia2016}.} \cite{AndrewsLu} propose a family of selection criteria by adding appropriate penalty and ``bonus'' terms to the J-test statistic, yielding analogues of AIC, BIC, and the Hannan-Quinn information criterion.
\cite{HongPrestonShum} extend this idea to generalized empirical likelihood (GEL). 
The principal goal of both papers is consistent selection: they state conditions under which the correct model and all correct moment conditions are chosen in the limit. 
As a refinement to this approach, \cite{LaiSmallLiu} suggest a two-step procedure: first consistently eliminate incorrect models using an empirical log-likelihood ratio criterion, and then select from the remaining models using a bootstrap covariance matrix estimator. 
The point of the second step is to address a shortcoming in the standard limit theory. 
While first-order asymptotic efficiency requires that we use all available correctly specified moment conditions, this can lead to a deterioration in finite sample performance if some conditions are only weakly informative.
\cite{HallPeixe2003} make a similar point about the dangers of including ``redundant'' moment conditions while \cite{Caner2009} proposes a lasso-type GMM estimator to consistently remove redundant parameters.

In contrast to these suggestions, the GFIC does not aim to identify the correct model and moment conditions: its goal is a low MSE estimate of a quantity of interest, even if this entails using a specification that is not exactly correct.  
As such, the GFIC is an ``efficient'' rather than a consistent selection criterion.
There is an unavoidable trade-off between consistent selection and estimators with desirable risk properties \citep{Yang2005}.
Indeed, the worst-case risk of any consistent selection procedure is \emph{unbounded} as sample size tends to infinity \citep{LeebPoetscher2008}.
In this sense, the fact that the GFIC is not consistent is a benefit rather than a liability.
As we show in simulations below, its worst-case performance is much better than that of competing selection procedures.

Although not strictly a selection procedure, the combined moments (CM) estimator of \cite{JudgeMittelhammer} takes a similar perspective to that of the GFIC, emphasizing that incoporating the information from an incorrect specification could lead to a favorable bias-variance tradeoff under the right circumstances. 
Unlike the GFIC, however, the CM estimator is not targeted to a particular research goal.
A key point of the GFIC is that two researchers using the same dataset but interested in different target parameters may find it optimal, in a minimum MSE sense, to choose different model specifications.  
We explore this idea further in our dynamic panel example below.

The remainder of this paper is organized as follows. Section \ref{sec:asymp} derives the asymptotic distribution of GMM estimators under locally mis-specified moment conditions and parameter restrictions. 
Section \ref{sec:GFIC} uses this information to calculate the AMSE of a user-specified target parameter and provides asymptotically unbiased estimators of the required bias parameters, yielding the GFIC. 
Section \ref{sec:avg} extends the results on averaging estimators and post-selection inference from \cite{DiTraglia2016} to the more general setting of this paper.
Section \ref{sec:Dpanel} specializes the GFIC to a dynamic panel example, and Section \ref{sec:Dpanel_sim} presents simulation results.
Finally, Section \ref{sec:cigarettes} presents our empirical example and  Section \ref{sec:conclude} concludes.  
Proofs and supplementary simulation results appear in the Appendix.
Further examples and simulation results appear in Online Appendices \ref{sec:additional} and \ref{sec:simulation_supplement}.

\section{Asymptotic Framework}
\label{sec:asymp}
Let $f(\cdot, \cdot)$ be a $(p+q)$-vector of moment functions of a random vector $Z$ and an $(r+s)$-dimensional parameter vector $\beta$.
To represent moment selection, we partition the moment functions according to $f(\cdot,\cdot) = \left(g(\cdot, \cdot)', h(\cdot, \cdot)'\right)'$ where $g(\cdot, \cdot)$ and $h(\cdot, \cdot)$ are $p$- and $q$-vectors.
The moment condition associated with $g(\cdot, \cdot)$ is assumed to be correct, while that associated with $h(\cdot,\cdot)$ is locally mis-specified.  
The moment selection problem is to choose which, if any, of the elements of $h$ to use in estimation. 
To represent model selection, we partition the full parameter vector according to $\beta = \left(\theta', \gamma'\right)'$, where $\theta$ is an $s$-vector and $\gamma$ an $r$-vector of parameters.
The model selection problem is to decide which if any of the elements of $\gamma$ to estimate, and which to set equal to the corresponding elements of $\gamma_0$, an $r$-vector of known constants.
The parameters contained in $\theta$ are those that we always estimate, the ``protected'' parameters.
Any specification that does not estimate the full parameter vector $\beta$ is locally mis-specified.
The precise form of the local mis-specification, over parameter restrictions and moment conditions, is as follows.

\begin{assump}[Local Mis-specification]
\label{assump:local}
Let $\{Z_{ni}\colon 1\leq i \leq n, n =1, 2, \hdots\}$ be an iid triangular array of random vectors defined on a probability space $(\Upsilon, \mathcal{F},P)$ satisfying
	\begin{enumerate}[(a)]
    \item $E[g(Z_{ni},\theta_0, \gamma_n)]=0$, with $\gamma_n = \gamma_0 + n^{-1/2} \delta$
    \item $E[h\left( Z_{ni}, \theta_0, \gamma_n,\right)] = \tau_n$, with $\tau_n = n^{-1/2} \tau$
		\item $\{f(Z_{ni}, \theta_0, \gamma_n)\colon 1\leq i \leq n, n = 1, 2, \hdots \}$ is uniformly integrable, and 
		\item $Z_{ni}\overset{d}{\rightarrow} Z_i$.
	\end{enumerate}
	where $\gamma_0$ is a (known) $r$-vector of parameter restrictions, $\delta$ an unknown $r$-vector of constants, and $\tau$ an unknown $q$-vector of constants.
\end{assump}

Assumption \ref{assump:local} specifies a triangular array data generating process in which the the true parameter vector $\beta_n = \left(\theta_0', \gamma_n'\right)'$, changes with sample size but converges to $\beta_0 = \left(\theta_0', \gamma_0'\right)'$ as $n\rightarrow \infty$.\footnote{For simplicity, and because it is the case for all examples we consider below, we assume that the triangular array from Assumption \ref{assump:local} is iid within each row. Note however, that the $Z_{ni}$ are not iid \emph{across} rows: $\gamma_n$ and $\tau_n$ change with $n$. As such, the triangular array machinery is still required to describe our results.}
Unless some elements of $\delta$ are zero, any estimator that restricts $\gamma$ is mis-specified for fixed $n$. 
In the limit, however, the restriction $\gamma = \gamma_0$ holds. 
Similarly, for any fixed sample size $n$, the expectation of $h$ evaluated at the true parameter value $\beta_n$ depends on the unknown constant vector $\tau$, but this source of mis-specification disappears in the limit. 
Thus, under Assumption \ref{assump:local}, only estimators that use moment conditions from $g$ to estimate the full parameter vector $\beta$ are correctly specified. 
In the limit, however, \emph{every} estimator is correctly specified, regardless of which elements of $\gamma$ it restricts and which elements of $h$ it includes. 
The purpose of local mis-specification is to ensure that squared asymptotic bias is of the same order as asymptotic variance: Assumption \ref{assump:local} is a device rather than literal description of real-world data. 
Note that, by Assumption \ref{assump:local}, the limiting random variable $Z_i$ satisfies the population moment condition $E[f\left(Z_i,\theta_0, \gamma_0\right)]=0$. 
Since the $Z_i$ are assumed to have a common marginal law, we will use the shorthand $Z$ for $Z_i$ throughout.
Accordingly, define:
		\begin{equation}F  =  \left[\begin{array}{cc}  \nabla_{\theta'}g\left(Z, \theta_0, \gamma_0\right) &   \nabla_{\gamma'}g\left(Z,\gamma_0, \theta_0\right) \\
 \nabla_{\theta'}h\left(Z,\gamma_0, \theta_0\right)  &   \nabla_{\gamma'}h\left(Z,\gamma_0, \gamma_0\right)  
		 \end{array}\right] = \left[\begin{array}{cc}F_\theta & F_\gamma \end{array}\right] = 
		  \left[\begin{array}{cc} G_\theta & G_\gamma \\
					 H_\theta& H_\gamma
		 \end{array}\right] = \left[\begin{array}{c} G\\ H \end{array}\right]
	\end{equation}
along with 
	\begin{equation}
		\Omega = Var\left[ \begin{array}{c} g(Z, \theta_0, \gamma_0) \\ h(Z, \theta_0, \gamma_0) \end{array}\right] =\left[ \begin{array}{cc}\Omega_{gg} & \Omega_{gh}\\ \Omega_{hg}& \Omega_{hh} \end{array} \right].
	\end{equation}
Each of these expressions involves the limiting random variable $Z$ rather than $Z_{ni}$. 
Thus, the corresponding expectations are taken with respect to a distribution for which all moment conditions have expectation zero  evaluated at $(\theta_0, \gamma_0)$.

Before defining the estimators under consideration, we require some further notation. Let $b$ be a \emph{model selection vector}, an $r$-vector of ones and zeros indicating which elements of $\gamma$ we have chosen to estimate. 
When $b = 1_r$, where $1_m$ represents an $m$-vector of ones, we estimate both $\theta$ and the full vector $\gamma$. 
When $b = 0_r$, where $0_m$ denotes an $m$-vector of zeros, we estimate only $\theta$, setting $\gamma=\gamma_0$. 
More generally, we estimate $|b|$ components of $\gamma$ and set the others equal to the corresponding elements of $\gamma_0$. 
Let $\gamma^{(b)}$ be the $|b|$-dimensional subvector of $\gamma$ corresponding to those elements selected for estimation. 
Similarly, let $\gamma^{(-b)}_0$ denote the $(r-|b|)$-dimensional subvector containing the values to which we set those components of $\gamma$ that are \emph{not} estimated.
Analogously, let $c=\left(c_g', c_h'\right)'$ be a \emph{moment selection vector}, a $(p+q)$-vector of ones and zeros indicating which of the moment conditions we have chosen to use in estimation. 
We denote by $|c|$ the total number of moment conditions used in estimation. 
Let $\mathcal{BC}$ denote the collection of all model and moment selection pairs $(b,c)$ under consideration.
To express moment and model selection in matrix form, we define the selection matrices $\Xi_b$ and $\Xi_c$.
Multiplying $\beta$ by the $(|b| + s)\times(r+s)$ \emph{model selection matrix} $\Xi_b$ extracts the elements corresponding to $\theta$ and the subset of $\gamma$ indicated by the model selection vector $b$. 
Thus $\Xi_b \beta = \left(\theta', \gamma^{(b)'} \right)'$.
Similarly, multiplying a vector by the $|c|\times(p+q)$ moment selection matrix $\Xi_c$ extracts the components corresponding to the moment conditions indicated by the moment selection vector $c$. 
To simplify the notation, we adopt the shorthand $F(b,c) = \Xi_c F \Xi_b'$ and $\Omega_c = \Xi_c \Omega \Xi_c'$ throughout.

To express the estimators themselves, define the sample analogue of the expectations in Assumption \ref{assump:local} as follows,
\begin{equation}
	f_n(\beta) = \frac{1}{n}\sum_{i=1}^n f(Z_{ni}, \theta, \gamma) = \left[\begin{array}{c}g_n(\beta)\\ h_n(\beta) \end{array} \right] = \left[\begin{array}{c} n^{-1}\sum_{i=1}^n g(Z_{ni}, \theta, \gamma)\\n^{-1}\sum_{i=1}^n h(Z_{ni}, \theta, \gamma) \end{array}\right]
\end{equation}
and let $\widetilde{W}$ be a $(p+q)\times(p+q)$ positive semi-definite weighting matrix with blocks $\widetilde{W}_{gg}, \widetilde{W}_{gh}, \widetilde{W}_{hg}$ and $\widetilde{W}_{hh}$, partitioned conformably to the partition of $f(Z,\beta)$ by $g(Z,\beta)$ and $h(Z,\beta)$. 
Each model and moment selection pair $(b,c)\in \mathcal{BC}$ defines a $(|b|+s)$-dimensional estimator $\widehat{\beta}(b,c)=(\widehat{\theta}(b,c)', \widehat{\gamma}^{(b)}(b,c)')'$ of $\beta^{(b)}= \left(\theta', \gamma^{(b)'} \right)'$ according to
	\begin{equation}
    \widehat{\beta}(b,c) = \underset{\beta^{(b)}\in \mathbf{B}^{(b)}} {\mbox{arg min}}\;\left[\Xi_c f_n\left(\beta^{(b)}, \gamma_0^{(-b)}\right)\right]'\left[\Xi_c \widetilde{W}\Xi_c' \right]\left[\Xi_c f_n\left(\beta^{(b)}, \gamma_0^{(-b)}\right)\right].
	\end{equation}

We now state a number of standard high-level regularity conditions that will be assumed throughout our derivations below.

\begin{assump}[High-level Regularity Conditions]
  \label{assump:high-level}
    \mbox{}
\begin{itemize}
\item[(a)] $\beta_0$ lies in the interior of $\Theta$, a compact set
\item[(b)] $\widetilde{W} \rightarrow_{p} W$, a positive definite matrix
\item[(c)] $W_c \Xi_c E[f(Z, \beta)] = 0$ if and only if $\beta = \beta_0$, where $W_c = \Xi_c W \Xi_c'$
\item[(d)] $E[f(Z,\beta)]$ is continuous on $\Theta$
\item[(e)] $sup_{\beta\in \Theta} ||f_n(\beta) - E[f(Z,\beta)]||\rightarrow_p 0$
\item[(f)] $f$ is Z-almost surely differentiable in an open neighborhood $\mathcal{B}$ of $\beta_0$
\item[(g)] $sup_{\beta \in \Theta} ||\nabla_{\beta} f_n(\beta) - F(\beta)|| \rightarrow_p 0 $
\item[(h)] $\sqrt{n}f_n(\beta_0) - \sqrt{n}E[f(Z,\beta_0)] \rightarrow_d \mathscr{N}$ where $\mathscr{N} \sim N_{p+q}(0,\Omega)$
\item[(i)] $F(b,c)' W_c F(b,c)$ is invertible, where $W_c = \Xi_c W \Xi_c'$
\end{itemize}
\end{assump}

  A particularly important special case is the estimator using only the moment conditions in $g$ to estimate the full parameter vector $\beta = \left(\theta', \gamma'\right)'$.
  We call this the \emph{valid} estimator and denote it by $\widehat{\beta}_v$.  
  Because it is assumed to be correctly specified both for finite $n$ and in the limit, the valid estimator contains the information we use to identify $\tau$ and $\delta$, and thus carry out moment and model selection. 
We assume that the valid estimator is identified.

\begin{assump}[GFIC Identification Condition]
  Let $\widehat{\beta}_v$ denote the GMM estimator for $\beta = \left(\theta', \gamma' \right)'$ based solely on the moment conditions contained in g:
\begin{equation}
  \widehat{\beta}_v = \left( \widehat{\theta}_{v}',\widehat{\gamma}_{v}'\right)' =\underset{\beta \in \mathbf{B}}{\mbox{arg min}}\; g_n(\beta)' \widetilde{W}_{gg} \; g_n(\beta).
\end{equation}
We call this the ``valid estimator'' and assume that it satisfies all the conditions of Assumption \ref{assump:high-level}.
In particular assume that $\widehat{\beta}_v$ is identified, which implies $p \geq r+s$.
\end{assump}

Because Assumption \ref{assump:local} ensures that they are correctly specified in the limit, \emph{all} candidate specifications $(b,c)\in \mathcal{BC}$ provide consistent estimators of $\theta_0$ under standard, high level regularity conditions (see Assumption \ref{assump:high-level}).
Essential differences arise, however, when we consider their respective asymptotic distributions. 
Under Assumption \ref{assump:local}, both $\delta$ and $\tau$ induce a bias term in the limiting distribution of $\sqrt{n}\left(\widehat{\beta}(b,c) - \beta_0^{(b)}\right)$. 

\begin{thm}[Asymptotic Distribution]
\label{thm:asymp}
Under Assumptions \ref{assump:local}--\ref{assump:high-level}
		\begin{equation}
		\sqrt{n}\left(\widehat{\beta}(b,c) - \beta_0^{(b)}\right) \overset{d}{\rightarrow} - K(b,c)\Xi_c \left(\mathscr{N}+ \left[ \begin{array}{c} 0\\ \tau\end{array}\right] - F_\gamma\delta\right)
	\end{equation}
where $\beta_0^{(b)'} = (\theta_0',\gamma_0^{(b)'})'$, $K(b,c) = \left[F(b,c)'W_c F(b,c)\right]^{-1} F(b,c)' W_c$
and $\mathscr{N} \sim \mbox{N}(0, \Omega)$ with $\mathscr{N} = (\mathscr{N}_g', \mathscr{N}_h')'$.
\end{thm}	
Because it employs the correct specification, the valid estimator of $\theta$ shows no asymptotic bias.
Moreover, the valid estimator of $\gamma$ has an asymptotic distribution that is centered around $\delta$, suggesting an estimator of this bias parameter.
\begin{cor}[Asymptotic Distribution of Valid Estimator]
\label{cor:valid}
Under Assumptions \ref{assump:local}--\ref{assump:high-level}
		$$\sqrt{n}\left( \widehat{\beta}_v - \beta_0 \right) = \sqrt{n}\left(\begin{array}{c} \widehat{\theta}_v - \theta_0\\ \widehat{\gamma}_v - \gamma_0\end{array} \right) \overset{d}{\rightarrow}  \left[\begin{array}{c} 0\\ \delta\end{array}\right] -K_v \mathscr{N}_g$$
where $K_v = \left[G'W_{gg}G\right]^{-1}G'W_{gg}$ and $W_{gg} = \mbox{plim }\widetilde{W}_{gg}$.
\end{cor}

\section{The GFIC}
\label{sec:GFIC}
The GFIC chooses among potentially incorrect moment conditions and parameter restrictions to minimize estimator AMSE for a scalar target parameter. 
Denote this target parameter by $\mu = \varphi(\theta, \gamma)$, where  $\varphi$ is a real-valued, almost surely continuous function of the underlying model parameters $\theta$ and $\gamma$.
Let $\mu_n = \varphi(\theta_0,\gamma_n)$ and define $\mu_0$ and $\widehat{\mu}(b,c)$  analogously.
By Theorem \ref{thm:asymp} and the delta method, we have the following result.
\begin{cor} Under the hypotheses of Theorem \ref{thm:asymp},
		$$\sqrt{n}\left(\widehat{\mu}(b,c) - \mu_0 \right) \overset{d}{\rightarrow} -\nabla_\beta\varphi_0'\Xi_b '  K(b,c)\Xi_c \left(\mathscr{N}+ \left[ \begin{array}{c} 0\\ \tau\end{array}\right] - F_\gamma\delta\right) $$
where $\varphi_0 = \varphi(\theta_0, \gamma_0)$.
\end{cor}
The true value of $\mu$, however,  is $\mu_n$ rather than $\mu_0$ under Assumption \ref{assump:local}.
Accordingly, to calculate AMSE we recenter the limit distribution as follows.
\begin{cor} 
\label{cor:mulimit}
Under the hypotheses of Theorem \ref{thm:asymp},
		$$\sqrt{n}\left(\widehat{\mu}(b,c) - \mu_n \right) \overset{d}{\rightarrow} -\nabla_\beta\varphi_0'\Xi_b '  K(b,c)\Xi_c \left(\mathscr{N}+ \left[ \begin{array}{c} 0\\ \tau\end{array}\right] - F_\gamma\delta\right) -\nabla_\gamma \varphi_0' \delta$$
where $\varphi_0 = \varphi(\theta_0, \gamma_0)$.
\end{cor}

We see that the limiting distribution of $\widehat{\mu}(b,c)$ is not, in general, centered around zero: both $\tau$ and $\delta$ induce an asymptotic bias. 
Note that, while $\tau$ enters the limit distribution only once, $\delta$ has two distinct effects. 
First, like $\tau$, it shifts the limit distribution of $\sqrt{n}f_n(\theta_0, \gamma_0)$ away from zero, thereby influencing the asymptotic behavior of $\sqrt{n}\left(\widehat{\mu}(b,c) - \mu_0 \right)$. 
Second, unless the derivative of $\varphi$ with respect to $\gamma$ is zero at $(\theta_0,\gamma_0)$, $\delta$ induces a second source of bias when $\widehat{\mu}(b,c)$ is recentered around $\mu_n$. 
Crucially, this second source of bias exactly cancels the asymptotic bias present in the limit distribution of $\widehat{\gamma}_v$. 
Thus, the valid estimator of $\mu$ is asymptotically unbiased and its AMSE equals its asymptotic variance.
\begin{cor}
\label{cor:muvalid}
Under the hypotheses of Theorem \ref{thm:asymp},
	$$\sqrt{n}\left( \widehat{\mu}_v - \mu_n\right) \overset{d}{\rightarrow} -\nabla_\beta \varphi(\theta_0, \gamma_0)' K_v\mathscr{N}_g$$
where $\widehat{\mu}_v = \varphi(\widehat{\theta}_v, \widehat{\gamma}_v)$. Thus, the valid estimator $\widehat{\mu}_v$ shows no asymptotic bias and has asymptotic variance $\nabla_\beta \varphi(\theta_0, \gamma_0)'K_v \Omega_{gg}K_v'\nabla_\beta \varphi(\theta_0,\gamma_0)$.
\end{cor}
 
Using Corollary \ref{cor:mulimit}, the AMSE of $\widehat{\mu}(b,c)$ is as follows,
	\begin{align}
	\label{eq:AMSE}
		\mbox{AMSE}\left(\widehat{\mu}(b,c)\right) &= \mbox{AVAR}\left(\widehat{\mu}\left(b,c\right)\right)  + \mbox{BIAS}\left(\widehat{\mu}\left(b,c\right)\right)^2\\
		\mbox{AVAR}\left(\widehat{\mu}\left(b,c\right)\right) &= \nabla_\beta\varphi_0'\Xi_b '  K(b,c)\Omega_c K(b,c)'\Xi_b\nabla_\beta\varphi_0\\
		\mbox{BIAS}\left(\widehat{\mu}\left(b,c\right)\right) &= -\nabla_{\beta} \varphi_0' M(b,c)\left[\begin{array}{c} \delta \\ \tau\end{array} \right]
		\label{eq:bias}
\end{align}
where
\begin{equation}
	M(b,c) = \Xi_b'K(b,c) \Xi_c \left[\begin{array}{cc} -G_\gamma & 0 \\ -H_\gamma & I \end{array} \right] +
  \left[\begin{array}{ll} 
  0_{p\times r} & 0_{s\times q} \\
    I_r & 0_{r\times q}  
\end{array} 
  \right]
\end{equation}

The idea behind the GFIC is to construct an estimate $\widehat{\mbox{AMSE}}\left(\widehat{\mu}(b,c)\right)$ and choose the specification $(b^*,c^*)\in\mathcal{BC}$ that makes this quantity as small as possible. 
As a side-effect of the consistency of the estimators $\widehat{\beta}(b,c)$, the usual sample analogues provide consistent estimators of $K(b,c)$ and $F_{\gamma}' = (G_\gamma', H_\gamma')$ under Assumption \ref{assump:local}, and $\varphi(\widehat{\theta}_v,\gamma_0)$ is consistent for $\varphi_0$. 
Consistent estimators of $\Omega$ are also readily available under local mis-specification although the best choice may depend on the situation.\footnote{We discuss this in more detail for our dynamic panel example in Section \ref{sec:Dpanel} below.}
Since $\gamma_0$ is known, as are $\Xi_b$ and $\Xi_c$, only $\delta$ and $\tau$ remain to be estimated. 
Unfortunately, neither of these quantities is consistently estimable under local mis-specification. 
Intuitively, the data become less and less informative about $\tau$ and $\delta$ as the sample size increases since each term is divided by $\sqrt{n}$. 
Multiplying through by $\sqrt{n}$ counteracts this effect, but also stabilizes the variance of our estimators. 
Hence, the best we can do is to construct \emph{asymptotically unbiased} estimators of $\tau$ and $\delta$. 
Corollary \ref{cor:deltahat} provides the required estimator for $\delta$, namely $\widehat{\delta} = \sqrt{n}\left(\widehat{\gamma}_v - \gamma_0\right)$, while Lemma \ref{lem:tauhatasymp} provides an asymptotically unbiased estimator of $\tau$ by plugging $\widehat{\beta}_v$ into the sample analogue of the $h$-block of moment conditions.

\begin{cor} 
\label{cor:deltahat}
Under the hypotheses of Theorem \ref{thm:asymp},
	$\widehat{\delta} = \sqrt{n}\left(\widehat{\gamma}_v - \gamma_0\right) \overset{d}{\rightarrow} \delta - K_{v}^{\gamma} \mathscr{N}_g$
where 
$K_v = \left[G'W_{gg}G\right]^{-1}G'W_{gg} = \left(K_{v}^{\theta'}, K_{v}^{\gamma'}\right)'$, so $\widehat{\delta}$ is an asymptotically unbiased estimator of $\delta$.
\end{cor} 

\begin{lem}
\label{lem:tauhatasymp}
Under the hypotheses of Theorem \ref{thm:asymp},
	$$\widehat{\tau} = \sqrt{n}h_n\left(\widehat{\beta}_v\right) \overset{d}{\rightarrow} \tau - HK_v \mathscr{N}_g + \mathscr{N}_h$$
where $K_v = \left[G'W_{gg}G\right]^{-1}G'W_{gg}$. Hence, $\widehat{\tau}$ is an asymptotically unbiased estimator of $\tau$.
\end{lem}

Combining Corollary \ref{cor:deltahat} and Lemma \ref{lem:tauhatasymp}, gives the joint distribution of $\widehat{\delta}$ and $\widehat{\tau}$.

\begin{thm}
\label{thm:jointbias}
Under the hypotheses of Theorem \ref{thm:asymp},
	\[\left[\begin{array}{c}\widehat{\delta}\\ \widehat{\tau}\end{array}\right] = \sqrt{n}\left[\begin{array}{c}\left(\widehat{\gamma}_v-\gamma_0\right)\\h_n(\widehat{\beta}_v)\end{array}\right]
  \overset{d}{\rightarrow} \left[\begin{array}{c}\delta\\ \tau\end{array} \right] +\Psi \mathscr{N}, \quad
  \Psi = \left[\begin{array}{cc} -K_{v}^\gamma&\mathbf{0} \\ -HK_v&I\end{array}\right]
\]
	where $K_v = \left[G'W_{gg}G\right]^{-1}G'W_{gg}$  is partitioned according to $K_v' = (K_v^{\theta'}, K_v^{\gamma'})$.
\end{thm}
Now, we see immediately from Equation \ref{eq:bias} that
$$\mbox{BIAS}\left(\widehat{\mu}\left(b,c\right)\right)^2 = \nabla_\beta \varphi_0' M(b,c) \left[\begin{array}{cc}  \delta \delta'& \delta \tau'\\ \tau \delta'& \tau \tau'\end{array}\right] M(b,c)' \nabla_\beta \varphi_0$$
Thus, the bias parameters $\tau$ and $\delta$ enter the AMSE expression in Equation \ref{eq:AMSE} as outer products: $\tau\tau'$, $\delta\delta'$ and $\tau\delta'$.
Although $\widehat{\tau}$ and $\widehat{\delta}$ are asymptotically unbiased estimators of $\tau$ and $\delta$, it does \emph{not} follow that $\widehat{\tau}\widehat{\tau}'$, $\widehat{\delta}\widehat{\delta}'$ and $\widehat{\tau}\widehat{\delta}'$ are  asymptotically unbiased estimators of $\tau\tau'$, $\delta\delta'$, and $\tau\delta'$. 
The following result shows how to adjust these quantities to provide the required asymptotically unbiased estimates. 

\begin{cor}
\label{cor:biasestimators}
Suppose that $\widehat{\Psi}$ and $\widehat{\Omega}$ are consistent estimators of $\Psi$ and $\Omega$. Then, $\widehat{B}$ is an asymptotically unbiased estimator of $B$, where
\[
  \widehat{B} = \left[\begin{array}{cc}  \widehat{\delta} \widehat{\delta}'& \widehat{\delta} \widehat{\tau}'\\ \widehat{\tau} \widehat{\delta}'& \widehat{\tau} \widehat{\tau}'\end{array}\right] - \widehat{\Psi} \widehat{\Omega} \widehat{\Psi}', \quad B = 
\left[\begin{array}{cc}  \delta \delta'& \delta \tau'\\ \tau \delta'& \tau \tau'\end{array}\right].
\]
\end{cor}
Combining Corollary \ref{cor:biasestimators} with consistent estimates of the remaining quantities yields the GFIC, an asymptotically unbiased estimator of the AMSE of our estimator of a target parameter $\mu$ under each specification $(b,c)\in \mathcal{BC}$
\begin{equation}
\mbox{GFIC}(b,c) =\nabla_\beta \widehat{\varphi}_0' \left[\Xi_b' \widehat{K}(b,c)\widehat{\Omega}_c \widehat{K}(b,c)'\Xi_b +  \widehat{M}(b,c) \;\widehat{B} \; \widehat{M}(b,c)'\right]\nabla_\beta \widehat{\varphi}_0.
\label{eq:GFIC}
\end{equation}
We choose the specification $(b^*,c^*)$ that minimizes the GFIC over the candidate set $\mathcal{BC}$.

\section{Averaging and Post-Selection Inference}
\label{sec:avg}

While we are primarily concerned in this paper with the mean-squared error performance of our proposed selection techniques, it is important to have tools for carrying out  inference post-selection.
In this section we briefly present results that can be used to carry out valid inference for a range of model averaging and post-selection estimators, including the GFIC.\footnote{We direct the reader to \cite{DiTraglia2016} and the references contained therein for a more detailed discussion of inference post-selection.}

The GFIC is an efficient rather than consistent selection criterion: it aims to estimate a particular target parameter with minimum AMSE rather than selecting the correct specification with probability approaching one in the limit.
As pointed out by \cite{Yang2005}, among others, there is an unavoidable trade-off between consistent selection and desirable risk properties.
Faced with this dilemma, the GFIC sacrifices consistency in the interest of low AMSE.
Because it is not a consistent criterion, the GFIC remains random \emph{even in the limit}.
We can see this from Equation \ref{eq:GFIC} in Section \ref{sec:GFIC} and Corollary \ref{cor:biasestimators}.
While the quantities $\nabla_\beta \widehat{\varphi}_0$, $\widehat{K}(b,c)$, and $\widehat{\Omega}_c$ are consistent estimators of their population counterparts, $\widehat{B}$ is only an asymptotically unbiased estimator of $B$ and thus has a limiting distribution.
In particular $\widehat{B} \rightarrow_d \mathscr{B}(\mathscr{N}, \delta, \tau)$ where
\begin{equation}
  \mathscr{B}(\mathscr{N}, \delta, \tau) = 
  \left(\left[
  \begin{array}{c}
    \delta \\ \tau
  \end{array}
\right] + \Psi \mathscr{N}\right)
  \left(\left[
  \begin{array}{c}
    \delta \\ \tau
  \end{array}
\right] + \Psi \mathscr{N}\right)' - \Psi \Omega \Psi
\end{equation}
Accordingly, to carry out inference post-GFIC, we need a limiting theory that is rich enough to accommodate \emph{randomly-weighted} averages of the candidate estimators $\widehat{\mu}(b,c)$.
To this end, consider an estimator of the form $\widehat{\mu} = \sum_{(b,c) \in \mathcal{BC}} \widehat{\omega}(b,c) \widehat{\mu}(b,c)$
where $\widehat{\mu}(b,c)$ denotes the target parameter under the moment conditions and parameter restrictions indexed by $(b,c)$, $\mathcal{BC}$ denotes the full set of candidate specifications, and $\widehat{\omega}(b,c)$ denotes a collection of data-dependent weights.
These could be zero-one weights correponding to a moment or model selction criterion, e.g.\ select the estimator that minimizes GFIC, or model averaging weights.\footnote{For an example that averages over fixed and random effects estimators, see Section \ref{sec:REvsFE}.}
We impose the following mild restrictions on the weights $\widehat{\omega}$.

\begin{assump}[Conditions on the Weights]\mbox{}
	\begin{enumerate}[(a)] 
		\item $\sum_{(b,c) \in \mathcal{BC}} \widehat{\omega}(b,c) = 1$, almost surely
    \item For each $(b,c) \in \mathcal{BC}$, $\widehat{\omega}(b,c) \overset{d}{\rightarrow} \psi(\mathscr{N}, \delta, \tau|b,c)$,  a function of $\mathscr{N}$, $\delta$, $\tau$, and consistently estimable constants with at most countably many discontinuities.
	\end{enumerate}
\label{assump:weight}
\end{assump}

Under the preceding conditions, we can derive the limit distribution of $\widehat{\mu}$ shown in the following Corollary.

\begin{cor}[Limit Distribution of Averaging Estimators]
  Under Assumption \ref{assump:weight} and the hypotheses of Theorem \ref{thm:asymp},  
	$\sqrt{n}\left(\widehat{\mu} - \mu_n\right) \overset{d}{\rightarrow} \Lambda(\delta,\tau)$
where
	\begin{equation}
		\Lambda(\delta,\tau) = -\nabla_\beta\varphi_0' \sum_{(b,c) \in \mathcal{BC}} \psi(\mathscr{N},\delta, \tau|b,c) \left\{\Xi_b' K(b,c) \Xi_c \mathscr{N} + M(b,c)  \left[\begin{array}{c}\delta \\ \tau \end{array} \right]\right\}
	\end{equation}
  \label{cor:avg}
\end{cor}
Note that the limit distribution from the preceding corollary is highly non-normal: it is a \emph{randomly} weighted average of a normal random vector, $\mathscr{N}$.
To tabulate this distribution for the purposes of inference, we will in general need to resort to simulation.
If $\tau$ and $\delta$ were known, the story would end here.
In this case we could simply substitute consistent estimators of $K$ and $M$ and then repeatedly draw $\mathscr{N} \sim N(0, \widehat{\Omega})$, where $\widehat{\Omega}$ is a consistent estimator of $\Omega$, to tabulate the distribution of $\Lambda$ to arbitrary precision as follows.

\begin{alg}[Approximating Quantiles of $\Lambda(\delta,\tau)$]
\mbox{}
		\begin{enumerate}
    \item Generate $J$ independent draws $\mathscr{N}_j \sim N(0, \widehat{\Omega})$
			\item Set $\Lambda_j(\delta, \tau)= -\nabla_\beta\widehat{\varphi}_0' \sum_{(b,c) \in \mathcal{BC}} \widehat{\psi}(\mathscr{N}_j,\delta, \tau|b,c) \left\{\Xi_b' \widehat{K}(b,c) \Xi_c \mathscr{N}_j + \widehat{M}(b,c)  \left[\begin{array}{c}\delta \\ \tau \end{array} \right]\right\}$
			\item Using the $\Lambda_j(\delta, \tau)$, find $\widehat{a}(\delta,\tau)$, $\widehat{b}(\delta, \tau)$ so that
		$P\left\{ \widehat{a}(\delta,\tau) \leq\Lambda(\delta,\tau)\leq \widehat{b}(\delta,\tau) \right\} = 1 - \alpha$.
		\end{enumerate}
    \label{alg:fixed_tau_delta}
\end{alg}

Unfortunately, no consistent estimators of $\tau$ or $\delta$ exist: all we have at our disposal are asymptotically unbiased estimators.
The following ``1-step'' confidence interval is constructed by substituting these into Algorithm \ref{alg:fixed_tau_delta}.

\begin{alg}[1-Step Confidence Interval] 
  \label{alg:1step}
  Carry out of Algorithm \ref{alg:fixed_tau_delta} with $\tau$ and $\delta$ set equal to the estimators $\widehat{\tau}$ and $\widehat{\delta}$ from Theorem \ref{thm:jointbias} to calculate $\widehat{a}(\widehat{\delta}, \widehat{\tau})$ and $\widehat{b}(\widehat{\delta}, \widehat{\tau})$.
  Then set $\mbox{CI}_{1}(\alpha) = \left[ \widehat{\mu} - \widehat{b}(\widehat{\delta}, \widehat{\tau})/\sqrt{n}, \quad \widehat{\mu} - \widehat{a}(\widehat{\delta}, \widehat{\tau})/\sqrt{n} \right]$.
\end{alg}

The 1-Step interval defined in Algorithm \ref{alg:1step} is conceptually simple, easy to compute, and can perform well in practice.\footnote{For more discussion on this point, see \cite{DiTraglia2016}.}
But as it fails to account for sampling uncertainty in $\widehat{\tau}$, $\mbox{CI}_1$ does \emph{not} necessarily yield asymptotically valid inference for $\mu$.
Fully valid inference requires the addition of a second step to the algorithm and comes at a cost: conservative rather than exact inference.
In particular, the two-step procedure described in the following algorithm is guaranteed to yield an interval with asymptotic coverage probability of \emph{no less than} $(1- \alpha_1 - \alpha_2)\times 100\%$.

\begin{alg}[2-Step Confidence Interval for $\widehat{\mu}$]
\mbox{}
\begin{enumerate}
  \item Construct $\mathscr{R}$, a $(1-\alpha_1)\times 100\%$ joint confidence region for $(\delta,\tau)$ using Theorem \ref{thm:jointbias}.
  \item For each $(\delta^*,\tau^*)\in \mathscr{R}$ carry out Algorithm \ref{alg:fixed_tau_delta} with $\alpha = \alpha_2$ yielding a $(1 - \alpha_2) \times 100\%$ confidence interval $\left[ \widehat{a}(\delta^*,\tau^*),\; \widehat{b}(\delta^*,\tau^*) \right]$ for $\Lambda(\delta^*, \tau^*)$. 
  \item Set $\displaystyle \widehat{a}_{min} = \min_{(\delta^*,\tau^*)\in \mathscr{R}} \widehat{a}(\delta^*,\tau^*)$ and $\displaystyle \widehat{b}_{max} = \max_{(\delta^*,\tau^*)\in \mathscr{R}} \widehat{b}(\delta^*, \tau^*)$.
  \item Construct the interval $\mbox{CI}_2(\alpha_1, \alpha_2) = \left[ \widehat{\mu} - \widehat{b}_{max}/\sqrt{n}, \; \widehat{\mu} - \widehat{a}_{min}/\sqrt{n} \right]$.
\end{enumerate}
\label{alg:2step}
\end{alg}

\begin{thm}[2-Step Confidence Interval for $\widehat{\mu}$]
\label{thm:sim}
Let $\nabla_{\beta}\widehat{\varphi}_0$, $\widehat{\psi}(\cdot|b,c)$, $\widehat{K}(b,c)$ and $\widehat{M}(b,c)$ be consistent estimators of $\nabla_\beta \varphi_0$, $\psi(\cdot|b,c)$, $K(b,c)$ and $M(b,c)$ and let $R$ be a $(1-\alpha_1)\times 100\%$ confidence region for $(\delta,\tau)$ constructed from Theorem \ref{thm:jointbias}.
Then $CI_2(\alpha_1, \alpha_2)$, defined in Algorithm \ref{alg:2step} has asymptotic coverage probability no less than $1-\left( \alpha_1 + \alpha_2 \right)$ as $J,n\rightarrow \infty$. 
\end{thm}

\section{Dynamic Panel Example}
\label{sec:Dpanel}
We now specialize the GFIC to a dynamic panel model of the form 
\begin{equation}
  y_{it} = \theta x_{it} + \gamma_1 y_{it-1} + \cdots + \gamma_k y_{it-k} + \eta_i + v_{it}
  \label{eq:truepanel}
\end{equation}
where $i = 1, \hdots, n$ indexes individuals and $t=1, \hdots, T$  indexes time periods. 
For simplicity, and without loss of generality, we suppose that there are no exogenous time-varying regressors and that all random variables are mean zero.\footnote{Alternatively, we can simply de-mean and project out any time-varying exogenous covariates after taking first-differences.} 
The unobserved error $\eta_i$ is a correlated individual effect: $\sigma_{x\eta}\equiv \mathbb{E}\left[ x_{it}\eta_i \right]$ may not equal zero. 
The endogenous regressor $x_{it}$ is assumed to be predetermined but not necessarily strictly exogenous: $\mathbb{E}[x_{it} v_{is}]=0$ for all $s \geq t$ but may be nonzero for $s < t$.  
We assume throughout that $y_{it}$ is stationary, which requires both $x_{it}$ and $u_{it}$ to be stationary and $|\boldsymbol{\gamma}| < 1$ where $\boldsymbol{\gamma} = (\gamma_1, \dots, \gamma_k)'$.
Our goal is to estimate one of the following two target parameters with minimum MSE: 
\begin{equation}
  \mu_{\text{SR}} \equiv \theta, \quad \quad 
  \mu_{\text{LR}} \equiv \theta / \left[1- (\gamma_1 + \cdots + \gamma_k)\right]
  \label{eq:paneltarget}
\end{equation}
where $\mu_{\text{SR}}$ denotes the short-run effect and $\mu_{\text{LR}}$ the long-run effect of $x$ on $y$.

The question is which assumptions to use in estimation.
Naturally, the answer may depend on whether our target is $\mu_{SR}$ or $\mu_{LR}$.
Our first decision is what assumption to impose on the relationship between $x_{it}$ and $v_{it}$.
This is the \emph{moment selection} decision.
We assumed above that $x$ is predetermined.
Imposing the stronger assumption of strict exogeneity gives us more and stronger moment conditions, but using these in estimation introduces a bias if $x$ is not in fact strictly exogenous.
Our second decision is how many lags of $y$ to use in estimation.
This is the \emph{model selection} decision.
The true model contains $k$ lags of $y$.
If we estimate only $r < k$ lags we not only have more degrees of freedom but more observations: every additional lag of $y$ requires us to drop one time period from estimation. 
In the short panel datasets common in microeconomic applications, losing even one additional time period can represent a substantial loss of information.
At the same time, unless $\gamma_{r+1} = \cdots = \gamma_k = 0$, failing to include all $k$ lags in the model introduces a bias.

To eliminate the individual effects $\eta_i$ we work in first differences.
Defining $\Delta$ in the usual way, so that $\Delta y_{it} = y_{it} - y_{it-1}$ and so on, we can write Equation \ref{eq:truepanel} as
\begin{align}
  \Delta y_{it} = \theta \Delta x_{it} + \gamma_1 \Delta y_{it-1} + \cdots + \gamma_k \Delta y_{it-k}  + \Delta v_{it}.
  \label{eq:truepaneldiff}
\end{align}
For simplicity and to avoid many instruments problems -- see e.g.\ \cite{Roodman} -- we focus here on estimation using the instrument sets
\begin{align}
  \mathbf{z}'_{it}(\ell, \text{P}) &\equiv \left[
  \begin{array}{cccc}
    y_{it-2} & \cdots & y_{it-(\ell + 1)} & x_{it-1}
  \end{array}
\right] & 
\mathbf{z}'_{it}(\ell,\text{S}) &\equiv \left[
\begin{array}{cc}
  \mathbf{z}_{it}'(\ell,\text{P}) & x_{it}
\end{array}
\right]
\label{eq:Zdpanel}
\end{align}
similar to \cite{AndersonHsiao}.
Modulo a change in notation, one could just as easily proceed using the instrument sets suggested by \cite{ArellanoBond}.
We use $\ell$ as a placeholder for the lag length used in estimation.
If $\ell = 0$, $\mathbf{z}'_{it}(0,\text{P}) = x_{it-1}$ and $\mathbf{z}'_{it}(0,\text{S}) = (x_{it-1}, x_{it})$.
Given these instrument sets, we have $(\ell + 1)\times (T -\ell - 1)$ moment conditions if $x$ is assumed to be predetermined versus $(\ell + 2)\times (T - \ell - 1)$ if it is assumed to be strictly exogenous, corresponding to the instrument matrices
  $Z_i(\ell,\text{P}) = \mbox{diag}\left\{ \mathbf{z}'_{it}(\ell,\text{P})  \right\}_{t = \ell + 2}^T$ and  $Z_i(\ell,\text{S}) = \mbox{diag}\left\{\mathbf{z}'_{it}(\ell,\text{S}) \right\}_{t = \ell +2}^T$.
To abstract for a moment from the model selection decision, suppose that we estimate a model with the true lag length: $\ell = k$.
The only difference between the P and S sets of moment conditions is that the latter adds over-identifying information in the form of $E[x_{it}\Delta v_{it}]$.
If $x$ is strictly exogenous, this expectation equals zero, but if $x$ is only predetermined, then $E[x_{it}\Delta v_{it}] = -E[x_{it}v_{it-1}] \neq 0$ so the over-identifying moment condition is invalid.
Given our instrument sets, this is the only violation of strict exogeneity that is relevant for our moment selection decision so we take $E[x_{it}v_{it-1}] = -\tau/\sqrt{n}$.

In the examples and simulations described below we consider two-stage least squares (TSLS) estimation of $\mu_{SR}$ and $\mu_{LR}$ using the instruments defined in Equation \ref{eq:Zdpanel}. 
Without loss of generality, we select between two lag length specifications: the first is correct, $\ell = k$, and the second includes $m$ lags too few: $\ell = r$ where $r = k-m$.
Accordingly, we make the coefficients associated with the $(r+1)$\textsuperscript{th}, $\ldots, k$\textsuperscript{th} lags local to zero.
Let $\boldsymbol{\gamma}' = (\gamma_1, \cdots, \gamma_{k-1}, \gamma_{k})$ denote the full vector of lag coefficients and $\boldsymbol{\gamma}_{r}' = (\gamma_1, \cdots, \gamma_{r})$ denote the first $r = k-m$ lag coefficients.
Then, the true parameter vector is $\beta_n = (\theta, \boldsymbol{\gamma}'_{r}, \boldsymbol{\delta}'/\sqrt{n})'$ which becomes, in the limit, $\beta = (\theta, \boldsymbol{\gamma}'_r, \boldsymbol{0}')'$. Both $\boldsymbol{\delta}$ and $\boldsymbol{0}$ are of length $m$. 
To indicate the subvector of $\beta$ that excludes the $(r+1)$\textsuperscript{th}, $\ldots, k$\textsuperscript{th} lag coefficients, let $\beta_{r} = (\theta, \boldsymbol{\gamma}_r')'$. 

Because the two lag specifications we consider use different time periods in estimation, we require some additional notation to make this clear. 
First let
$\Delta \mathbf{y}_{i} = [\Delta y_{i,k+2}, \cdots, \Delta y_{iT}]'$ and $\Delta \mathbf{y}^+_{i} = [\Delta y_{i,k+2-m}, \Delta y_{i,k+2-(m-1)}, \cdots, \Delta y_{iT}]'$ 
where the superscript ``+'' indicates the inclusion of $m$ additional time periods: $t = k+2-m, \ldots, k+1$.
Define $\Delta \mathbf{x}_i$, $\Delta \mathbf{x}_{i}^{+}$, $\Delta \mathbf{v}_i$, and $\Delta \mathbf{v}_{i}^{+}$ analogously.
Next, define $L^{r+1}\Delta \mathbf{y}_i^{+} = [\Delta y_{i1}, \Delta y_{i2}, \cdots, \Delta y_{iT-(r+1)}]'$ where $L^{r+1}$ denotes the element-wise application of the $(r+1)$\textsuperscript{th} order lag operator.
Note that the first element of $L^{r+1}\Delta \mathbf{y}_{i}^{+}$ is unobserved since $\Delta y_{i1} = y_{i1} - y_{i0}$ but $t=1$ is the first time period.
Now we define the matrices of regressors for the two specifications: 
\begin{align*}
  W_i^{+'}(r) &= \left[
  \begin{array}{ccccc}
    \Delta \mathbf{x}_i^+ & L \Delta \mathbf{y}_i^{+} &  L^2 \Delta \mathbf{y}_i^{+} & \cdots & L^{r}\Delta \mathbf{y}_i^{+} 
  \end{array}
\right]\\
  W_i'(k) &= \left[
  \begin{array}{cccccc}
    \Delta \mathbf{x}_i & L \Delta \mathbf{y}_i &  L^2 \Delta \mathbf{y}_i & \cdots & L^{k-1}\Delta \mathbf{y}_i & L^k\Delta \mathbf{y}_i 
  \end{array}
\right].
\end{align*}
Note that $W_i^{+}(r)$ contains $m$ more rows than $W_i(k)$ but $W_i(k)$ contains $m$ more columns than $W_i^{+}(r)$: removing the $(r+1)$\textsuperscript{th}, $\ldots, k$\textsuperscript{th} lags from the model by setting $\ell = r = k-m$ allows us to use $m$ additional time periods in estimation and reduces the number of regressors by $m$. 
Stacking over individuals, let $\Delta \mathbf{y} = [\Delta \mathbf{y}'_1 \cdots \Delta \mathbf{y}'_n]'$, $W_\ell = [W_1(\ell) \cdots W_n(\ell)]'$ and define $\Delta \mathbf{y}^{+}$ and $W_\ell^{+}$ analogously, where $\ell$ denotes the lag length used in estimation.
Finally, let $Z'(\ell,\cdot) = [Z'_1(\ell,\cdot) \cdots Z'_n(\ell,\cdot)]$ where $(\cdot)$ is $\text{P}$ or $\text{S}$ depending on the instrument set in use.
Using this notation, under local mis-specification the true model is
\begin{align}
  \Delta \mathbf{y} &= W(k)\beta_n + \Delta \mathbf{v} &  \Delta \mathbf{y}^{+} &= W(k)^{+}\beta_n + \Delta \mathbf{v}^+
\end{align}
Using the shorthand $\widehat{Q} \equiv n[W' Z(Z'Z)^{-1} Z'W]^{-1}W'Z(Z'Z)^{-1}$ our candidate estimators are
\begin{align}
  \widehat{\beta}(k,\cdot) &= \widehat{Q}(k,\cdot)\left[ \frac{Z'(k,\cdot)\Delta \mathbf{y}}{n} \right]& 
  \widehat{\beta}(r,\cdot) &= \widehat{Q}(r,\cdot)\left[ \frac{Z'(r,\cdot)\Delta \mathbf{y}^{+}}{n} \right]
  \label{eq:DpanelEstimators}
\end{align}
where $(\cdot)$ is either $\text{P}$ or $\text{S}$ depending on which instrument set is used and $r = k-m$, $m$ lags fewer than the true lag length $k$.
The following result describes the limit distribution of $\widehat{\beta}(k,\text{P})$, $\widehat{\beta}(k,\text{S})$, $\widehat{\beta}(r,\text{P})$, and $\widehat{\beta}(r,\text{S})$ which we will use to construct the GFIC.

\begin{thm}[Limit Distributions for Dynamic Panel Estimators]
  \label{thm:limitDpanel}
  Let $(y_{nit},x_{nit}, v_{nit})$ be a triangular array of random variables that is iid over $i$, stationary over $t$, and satisfies Equation \ref{eq:truepaneldiff} with $(\gamma_{k-m+1}, \ldots, \gamma_k)' = \boldsymbol{\delta} / \sqrt{n}$.
  Suppose further that $x_{it}$ is predetermined with respect to $v_{it}$ but not strictly exogenous: $E[x_{it}\Delta v_{it}] = \tau/\sqrt{n}$.
  Then, under standard regularity conditions,
  \begin{align*}
    \sqrt{n}\left[ \widehat{\beta}(k,\text{P})-\beta \right] &\rightarrow^d 
    \left[
    \begin{array}{ccc}
    0 & \mathbf{0}_{r}'& \boldsymbol{\delta}'
    \end{array}
  \right]' + 
    Q\left(k,\text{P} \right) \mbox{N}\left(\mathbf{0}, \mathcal{V}(k,\text{P})\right)  \\
    \sqrt{n}\left[ \widehat{\beta}(k,\text{S})-\beta \right] &\rightarrow^d 
    \left[
    \begin{array}{ccc}
    0 & \mathbf{0}_{r}'& \boldsymbol{\delta}'
    \end{array}
  \right]' + 
     Q\left(k,\text{S} \right) \left\{ \boldsymbol{\iota}_{T-(k +1)} \otimes \left[
    \begin{array}{c}
      \mathbf{0}_{k+1} \\ \tau
    \end{array}
  \right] + \mbox{N}\left(\mathbf{0}, \mathcal{V}(k,\text{S})\right)\right\}\\
    \sqrt{n}\left[ \widehat{\beta}(r,\text{P})- \beta_r \right] &\rightarrow^d Q(r,\text{P}) \left[\boldsymbol{\iota}_{T-(r+1)} \otimes  \boldsymbol{\psi}_{\text{P}}\, \boldsymbol{\delta} + \mbox{N}\left(\mathbf{0}, \mathcal{V}(r,\text{P}) \right) \right]\\
    \sqrt{n}\left[ \widehat{\beta}(r,\text{S})- \beta_r\right] &\rightarrow^d Q(r,\text{S}) \left[\boldsymbol{\iota}_{T-(r+1)} \otimes 
    \left(  \left[
  \begin{array}{c}
    \boldsymbol{\psi}_{\text{P}} \\ 
    \boldsymbol{\psi}_{\text{S}} 
\end{array}
\right] \boldsymbol{\delta} + \left[
\begin{array}{c}
  \mathbf{0}_{r+1} \\ \tau
\end{array}
\right]\right) + \mbox{N}\left( \mathbf{0}, \mathcal{V}\left(r,\text{S}\right) \right)\right]
  \end{align*}
  where $r = k - m$, $\beta' = (\theta, \gamma_1, \hdots, \gamma_{r}, \boldsymbol{0}_m')$, $\beta_r' = (\theta, \gamma_1, \hdots, \gamma_{r})$, $\mathcal{V}(k,\cdot) = \mbox{Var}\left[ Z_i(k,\cdot) \Delta \mathbf{v}_i  \right]$, $\mathcal{V}(r,\cdot) = \mbox{Var}\left[ Z_i(r,\cdot) \Delta \mathbf{v}^{+}_i  \right]$, $\widehat{Q}(\ell,\cdot) \rightarrow_p Q(\ell,\cdot)$, $\boldsymbol{\psi}_{\text{P}} = E[\textbf{z}_{it}(r,\text{P}) (\Delta y_{it -(r+1)}, \ldots, \Delta y_{it-k})]$, $\boldsymbol{\psi}_{\text{S}} = E[x_{it} (\Delta y_{it -(r+1)}, \ldots, \Delta y_{it-k})]$, $Z_i(\ell, \cdot)= \mbox{diag}\{\mathbf{z}_{it}'(\ell, \cdot)\}_{t=\ell+2}^T$, $\mathbf{z}_{it}(\ell,\cdot)$ is as in Equation \ref{eq:Zdpanel}, and $\boldsymbol{\iota}_{d}$ denotes a $d$-vector of ones.
\end{thm}

To operationalize the GFIC, we need to provide appropriate estimators of all quantities that appear in Theorem \ref{thm:limitDpanel}.
To estimate ${Q}(k,\text{P})$, ${Q}(k,\text{S})$, ${Q}(r,\text{P})$, and ${Q}(r,\text{S})$ we employ the usual sample analogues $\widehat{Q}(\cdot,\cdot)$ given above, which remain consistent under local mis-specification.
There are many consistent estimators for the variance matrices $\mathcal{V}(k,\text{P})$, $\mathcal{V}(k,\text{S})$, $\mathcal{V}(r,\text{P})$, $\mathcal{V}(r,\text{S})$ under local mis-specification.
In our simulations below, we employ the usual heteroskedasticity-consistent, panel-robust variance matrix estimator.
Because $E[\mathbf{z}_{it}(\ell,\text{S})\Delta v_{it}]\neq 0$, we center our estimators of $\mathcal{V}(\ell, \text{S})$ by subtracting the sample analogue of this expectation when calculating the sample variance.
We estimate $\boldsymbol{\psi}_{\text{P}}$ and $\boldsymbol{\psi}_{\text{S}}$ as follows
\[
  \widehat{\boldsymbol{\psi}}_{\text{P}}' = \frac{1}{nT_k}\begin{bmatrix}
  \sum_{t = k+2}^T \sum_{i = 1}^n \mathbf{z}_{it}(r,\text{P}) \Delta y_{it-(r+1)}\\
  \vdots \\
   \sum_{t = k+2}^T \sum_{i = 1}^n \mathbf{z}_{it}(r,\text{P}) \Delta y_{it-k}
\end{bmatrix},   \quad
\widehat{\boldsymbol{\psi}}_{\text{S}}' = \frac{1}{nT_k}\begin{bmatrix}
   \sum_{t = k+2}^T \sum_{i = 1}^n x_{it} \Delta y_{it-(r+1)}\\
   \vdots\\
    \sum_{t = k+2}^T \sum_{i = 1}^n x_{it} \Delta y_{it-k}
\end{bmatrix}  
\]
where $T_k = T-k-1$.
These estimates use our assumption of stationarity from above.
The only remaining quantities we need to construct the GFIC involve the bias parameters $\boldsymbol{\delta}$ and $\tau$. 
We can read off an asymptotically unbiased estimator of $\boldsymbol{\delta}$ directly from Theorem \ref{thm:limitDpanel}, namely $\widehat{\boldsymbol{\delta}} = \sqrt{n}\; (\widehat{\gamma}_{r+1}(k,\text{P}), \ldots, \widehat{\gamma}_k(k,\text{P}))'$ based on the instrument set that assumes only that $x$ is pre-determined rather than strictly exogenous.
To construct an asymptotically unbiased estimator of $\tau$, we use the residuals from the specification that uses \emph{both} the correct moment conditions and the correct lag specification, specifically
\begin{equation}
  \label{eq:DpanelTau}
  \widehat{\tau} = \left( \frac{\boldsymbol{\iota}_{T-k-1}'}{T - k - 1} \right) n^{-1/2} X' \left[\Delta \mathbf{y} - W(k)\widehat{\beta}(k,\text{P})  \right]
\end{equation}
where $X' = [X_1 \cdots X_n]$ and $X_i = \mbox{diag}\left\{ x_{it} \right\}_{t = k + 2}^{T}$.
The following result gives the joint limiting behavior of $\widehat{\boldsymbol{\delta}}$ and $\widehat{\tau}$, which we will use to construct the GFIC.

\begin{thm}[Joint Limit Distribution of $\widehat{\boldsymbol{\delta}}$ and $\widehat{\tau}$]
  \label{thm:DpanelJoint}
  Under the conditions of Theorem \ref{thm:limitDpanel},
  \[
    \left[
      \begin{array}{c} 
        \widehat{\boldsymbol{\delta}} - \boldsymbol{\delta} \\ \widehat{\tau} - \tau 
      \end{array} 
    \right] \overset{d}{\rightarrow} \Psi \mbox{N}\left(\mathbf{0}, \Pi\,\mathcal{V}\left(k,\text{S}\right)\,\Pi'\right)
  \]
  where $\widehat{\boldsymbol{\delta}} = \sqrt{n}[ \mathbf{e}_k' \,\widehat{\beta}(k,\text{P})]$, $\mathbf{e}_k = (0, \mathbf{0}_{k-m}', \boldsymbol{\iota}_m')'$,  $\widehat{\tau}$ is as defined in Equation \ref{eq:DpanelTau}, 
\[
  \Psi = \left[
  \begin{array}{cc}
    \displaystyle
    \mathbf{e}_k' Q(k,\text{P}) & \mathbf{0}'_{T-k-1}\\
    \left( \frac{\boldsymbol{\iota}'_{T-k-1}}{T-k-1} \right)  \left\{ \boldsymbol{\xi}' Q(k,\text{P}) \otimes \boldsymbol{\iota}'_{T-k-1} \right\}& \displaystyle \left(\frac{\boldsymbol{\iota}_{T-k-1}}{T-k-1}\right) 
  \end{array}
\right],
\]
  $\boldsymbol{\xi}' = E\left\{ x_{it} \left[
    \begin{array}{cccc}
       \Delta x_{it} & L \Delta y_{it} & \cdots & L^k \Delta y_{it}   \end{array} \right]\right\}$,
  the variance matrix $\mathcal{V}(k,\text{S})$ is as defined in Theorem \ref{thm:limitDpanel}, the permutation matrix $\Pi = \left[
  \begin{array}{cc}
    \Pi_1' & \Pi_2'
  \end{array}
\right]'$ with $\Pi_1 = I_{T-k-1} \otimes \left[
\begin{array}{cc}
  I_{k+1} & \mathbf{0}_{k+1}
\end{array}
\right]$ and $\Pi_2 = I_{T-k-1}\otimes \left[
\begin{array}{cc}
  \mathbf{0}_{k+1}' & 1
\end{array}
\right]$,
  $\boldsymbol{\iota}_{d}$ is a $d$-vector of ones and $I_d$ the $(d\times d)$ identity matrix.
\end{thm}

To provide asymptotically unbiased estimators of the quantities $\tau^2$, $\boldsymbol{\delta}\boldsymbol{\delta}'$ and $\boldsymbol{\delta}\tau$ that appear in the AMSE expressions for our estimators, we apply a bias correction to the asymptotically unbiased estimators of $\boldsymbol{\delta}$ and $\tau$ from Theorem \ref{thm:DpanelJoint}.

\begin{cor}
  Let $\widehat{\Psi}$ be a consistent estimator of $\Psi$, defined in Theorem \ref{thm:DpanelJoint}, and $\widehat{\mathcal{V}}(k,\text{S})$ be a consistent estimator of $\mathcal{V}(k,\text{S})$, defined in Theorem \ref{thm:limitDpanel}.
  Then, the elements of  
  \[
    \left[
    \begin{array}{cc}
      \widehat{\boldsymbol{\delta}} \widehat{\boldsymbol{\delta}}' & \widehat{\boldsymbol{\delta}} \widehat{\tau} \\
      \widehat{\tau} \widehat{\boldsymbol{\delta}}' & \widehat{\tau}^2
    \end{array}
  \right] - \widehat{\Psi}\, \Pi \, \widehat{V}(k,\text{S}) \, \Pi' \, \widehat{\Psi}'
  \]
  provide asymptotically unbiased estimators of of $\boldsymbol{\delta}\boldsymbol{\delta}'$, $\tau^2$ and $\boldsymbol{\delta}\tau$, where $\Pi$ is the permutation matrix defined in Theorem \ref{thm:DpanelJoint}.
\end{cor}

We have already discussed consistent estimation of $\widehat{\mathcal{V}}(k,\text{S})$.
Since $\Pi$ is a known permutation matrix, it remains only to propose a consistent estimator of $\Psi$. 
The matrix $\Psi$, in turn, depends only on $Q(k,\text{P})$, and $\boldsymbol{\xi}'$.
The sample analogue $\widehat{Q}(k,\text{P})$ is a consistent estimator for $Q(k,\text{P})$, as mentioned above, and
\begin{equation}
  \widehat{\boldsymbol{\xi}}' = \frac{1}{n(T - k - 1)} \sum_{t = k+2}^T \sum_{i=1}^n x_{it}\left[
  \begin{array}{cccc}
    \Delta x_{it} & L \Delta y_{it} & \cdots & L^{k} \Delta y_{it} 
  \end{array}
\right]
\end{equation}
is consistent for $\xi'$.
We now have all the quantities needed to construct the GFIC for $\mu_{SR}$, the short-run effect of $x$ on $y$.
Since $\mu_{SR} = \theta$, we can read off the AMSE expression for this parameter directly from Theorem \ref{thm:limitDpanel}.
For the long-run effect $\mu_{LR}$, however, we need to formally apply the Delta-method and account for the fact that the true value of $(\gamma_{r+1}, \ldots, \gamma_k)'$ is $\boldsymbol{\delta}/\sqrt{n}$.
Expressed as a function $\varphi$ of the underlying model parameters, 
\[
  \mu_{LR} = \varphi(\theta, \boldsymbol{\gamma}_r, \boldsymbol{\gamma}_{-r}) = \theta / \left[1 - \boldsymbol{\iota}_r' \boldsymbol{\gamma}_r - \boldsymbol{\iota}_m'\boldsymbol{\gamma}_{-r}\right]
\]
where we define $\boldsymbol{\gamma}_{-r} \equiv (\gamma_{r+1}, \ldots, \gamma_k)'$.
The derivatives of $\varphi$ are
\[
  \nabla \varphi \equiv \left[
  \begin{array}{ccc}
    \displaystyle\frac{\partial \varphi}{\partial \theta} & 
    \displaystyle\frac{\partial \varphi}{\partial \boldsymbol{\gamma}_r'} &
    \displaystyle\frac{\partial \varphi}{\partial \boldsymbol{\gamma}_{-r}'} 
  \end{array}
\right] = 
\left( \frac{1}{1 - \boldsymbol{\iota}_r' \boldsymbol{\gamma}_r - \boldsymbol{\iota}_m' \boldsymbol{\gamma}_{-r}} \right)^2 \left[
\begin{array}{ccc}
  \left( 1 - \boldsymbol{\iota}_r' \boldsymbol{\gamma}_r - \boldsymbol{\iota}_m'\boldsymbol{\gamma}_{-r} \right) & \theta \,\boldsymbol{\iota}_r' & \theta \, \boldsymbol{\iota}_m' 
\end{array}
\right].
\]
Using this notation, the limiting value of $\mu_{LR}$ is $\mu_{LR}^{0} =  \varphi(\theta, \boldsymbol{\gamma}_r, \boldsymbol{0}_m')$
while the true value is
$\mu_{LR}^{n} = \varphi(\theta, \boldsymbol{\gamma}_r, \boldsymbol{\delta}/\sqrt{n})$.
Similarly, the limiting value of $\nabla\varphi$ is $\nabla \varphi_0 = \nabla \varphi(\theta, \boldsymbol{\gamma}_r, \boldsymbol{0}_m')'$, obtained by putting zero in place of $\boldsymbol{\gamma}_{-r}$. 
We estimate this quantity consistently by plugging in the estimates from $\widehat{\beta}(k, \text{P})$.

\section{Simulation Study}
\label{sec:Dpanel_sim}
We now consider two simulation experiments based on section \ref{sec:Dpanel}, applying the GFIC to a dynamic panel model. 
For both experiments our data generating process is similar to that of \cite{AndrewsLu}, specifically
	\begin{equation}
	\label{eq:covar}
		\left[\begin{array}{c}
			x_{i}\\
			\eta_i\\
			v_{i}
  \end{array} \right]\sim \mbox{iid}\; N\left(\left[\begin{array}{c}\mathbf{0}_T\\ 0\\ \mathbf{0}_T \end{array}\right] ,\left[\begin{array}{ccc}
	 	 I_T & \sigma_{x\eta}\iota_T&\sigma_{xv}\Gamma_T \\
     \sigma_{x\eta}\iota_T'& 1&\mathbf{0}_T' \\
     \sigma_{xv}\Gamma_T'& \mathbf{0}_T&  I_T
	 \end{array}\right]\right), \,
		\Gamma_T = \left[\begin{array}{cc}
        \mathbf{0}_{T-1}' & 0\\
        I_{T-1} & \mathbf{0}_{T-1}
	 \end{array}\right].
	\end{equation}
where $\mathbf{0}_m$ denotes an $m$-vector of zeros, $I_m$ the $(m\times m)$ identity matrix, and $\iota_m$ an $m$-vector of ones.
Under this covariance matrix structure $\eta_i$ and $v_{i}$ are uncorrelated with each other, but both are correlated with $x_{i}$: $E[x_{it}\eta_i]=\sigma_{x\eta}$ and $x_{it}$ is predetermined but not strictly exogenous with respect to $v_{it}$. Specifically, $E[x_{it}v_{it-1}]=\sigma_{xv}$, while $E[x_{it}v_{is}]=0$ for $s\neq t-1$. 
We initialize the pre-sample observations of $y$ to zero, the mean of their stationary distribution, and generate the remaining time periods according to Equation \ref{eq:truepanel} with $\theta = 0.5$ and $\sigma_{x\eta} = 0.2$.
The true lag length differs in our two examples as does the target parameter, so we explain these features of the simulation designs below. 
Unlike \cite{AndrewsLu} we do not generate extra observations to keep the time dimension fixed across estimators with different lag specifications.
This is for two reasons. 
First, in real-world applications such additional observations would not be available. 
Second, we are explicitly interested in trading off the efficiency gain from including additional time periods in estimation against the bias that arises from estimating an incorrect lag specification. 

\subsection{Long-run versus Short-run Effects}
\label{sec:SRvsLR}
Consider two different researchers who happen to be working with the same panel dataset. 
One wishes to estimate the short-run effect of $x$ on $y$ while the other wishes to estimate the long-run effect.  
Should they use the same model specification?
We now present an example showing that the answer, in general, is no.
Suppose that the true model is
\[
  y_{it} = \theta x_{it} + \gamma_1 y_{it-1} + \gamma_2 y_{it-2}  + \eta_i + v_{it}
\]
where $i = 1, \dots, n = 250$ and $t = 1, \dots, T=5$ and the regressor, individual effect and error term are generated according to Equation \ref{eq:covar}, as described in the preceding section.
Our model selection decision in this example is whether to set $\gamma_2 = 0$ and estimate a specification with one lag only.
We denote this one-lag specification by $\mbox{L1}$ and the true specification, including both lags, by $\mbox{L2}$.
To focus on the model selection decision, we fix the instrument set in this experiment to $\mathbf{z}_{it}(\ell,\text{P})$, defined in Equation \ref{eq:Zdpanel}.
Because this instrument set is valid when $x$ is pre-determined, it does not introduce bias into our estimation.
Thus, bias only emerges if we estimate $\mbox{L1}$ when $\gamma_2\neq 0$.
Our simulation design takes $\theta = 0.5, \gamma_1 = 0.4, \sigma_{x\eta} = 0.2$, and $\sigma_{xv} = 0.1$ and varies $\gamma_2$ over the range $\{0.10, 0.11, \dots, 0.19, 0.20\}$. 

Table \ref{tab:MAD_SRvsLR} presents the results of the simulation, based on 1000 replications at each grid point.
Because they are based on \emph{ratios} of estimators of $\theta$ and $\gamma_1, \gamma_2$, estimators of the long-run effect may not have finite moments, making finite-sample MSE undefined.
The usual solution to this problem in simulation settings is to work with so-called ``trimmed'' MSE by discarding observations that fall outside, say, a range $[-M, M]$ before calculating MSE.\footnote{Note that \emph{asymptotic} MSE remains well-defined even for estimators that do not possess finite-sample moments so that GFIC comparisons remain meaningful. By taking the trimming constant $M$ to infinity, one can formalize the notion that asymptotic MSE comparisons can be used to ``stand in'' for finite-sample MSE even when the latter does not exist. For more details, See \cite{HansenShrink} and online appendix C of \cite{DiTraglia2016}.}
Because there is no clear way to set the trimming constant $M$, it can be difficult to interpret results based on trimmed MSE unless one considers a variety of values of $M$. 
To avoid this issue, Table \ref{tab:MAD_SRvsLR} reports simulation results for median absolute deviation (MAD).
Results for trimmed MSE with different choices of $M$ are similar and are available upon request.

\begin{table}[!hpt]
\centering
\small
\begin{tabular}{  c  c c  c  c c c  }
\hline
\hline
 & \multicolumn{3}{c}{Short-run Effect} & \multicolumn{3}{c}{Long-run Effect} \\
    $\gamma_2$ &      L2  &       L1   &    GFIC    & L2 &     L1 &   GFIC\\
    \hline
 0.10  &0.231&\bf{\color{blue}0.141}& 0.173& 0.801& \bf{\color{blue}0.582}& 0.688\\
  0.11 & 0.237& \bf{\color{blue}0.156}& 0.181& 0.834& \bf{\color{blue}0.633}& 0.716\\
 0.12   &0.240& \bf{\color{blue}0.174}& 0.193& 0.850& \bf{\color{blue}0.685}& 0.752\\
 0.13   &0.238& \bf{\color{blue}0.187}& 0.201& 0.870& \bf{\color{blue}0.729}& 0.787\\
 0.14 &0.220& \bf{\color{blue}0.198}& 0.203& 0.870& \bf{\color{blue}0.764}& 0.808\\
\bf{\color{red} 0.15}  & \bf{\color{blue}0.201}& 0.219& 0.211& 0.844& \bf{\color{blue}0.822}& 0.839\\
\bf{\color{red} 0.16}  & \bf{\color{blue}0.205}& 0.223& 0.210& 0.883& \bf{\color{blue}0.856}& 0.862\\
 0.17  &\bf{\color{blue}0.181}& 0.242& 0.204& \bf{\color{blue}0.860}& 0.911& 0.897\\
 0.18  & \bf{\color{blue}0.162}& 0.258& 0.189& \bf{\color{blue}0.835}& 0.959& 0.891\\
 0.19  &\bf{\color{blue}0.161}& 0.265& 0.181& \bf{\color{blue}0.866}& 0.997& 0.917\\
 0.20  &\bf{\color{blue}0.143}& 0.288& 0.162& \bf{\color{blue}0.858}& 1.054& 0.910 \\
\hline
 \hline
\end{tabular}
\caption{Comparisons of mean absolute deviation (MAD) for estimators of the Short-run and Long-run effects of $x$ on $y$ in the simulation experiment described in Section \ref{sec:SRvsLR}.
The columns labeled $\mbox{L1}$ and $\mbox{L2}$ give the MAD of estimators that fix the lag length to one and two, while the columns labeled GFIC give the MAD of an estimator that selects lag length via the GFIC.
Results are based on 1000 simulation replications from the DGP described in Section \ref{sec:Dpanel_sim} with $\gamma_1 = 0.4$, using the estimators described in Section \ref{sec:Dpanel} and the instrument set $\mathbf{z}_{it}(\ell, \text{P})$ from Equation \ref{eq:Zdpanel}.}
\label{tab:MAD_SRvsLR}
\end{table}		

The columns of Table \ref{tab:MAD_SRvsLR} labeled $\mbox{L1}$ and $\mbox{L2}$ give the MAD of estimators that fix the lag length to one and two, while those labeled GFIC give the MAD of an estimator that selects lag length via the GFIC.
Notice that throughout the table $\gamma_2 \neq 0$ so that $\mbox{L1}$ is \emph{mis-specified}.
Nonetheless, $\mbox{L1}$ yields lower MAD estimators of both the short-run and long-run effects when $\gamma_2$ is sufficiently small and the difference can be substantial.
When $\gamma_2 = 0.2$, for example, MAD for is 0.582 for the long-run effect estimator based on $\mbox{L1}$ versus 0.801 for that based on $\mbox{L2}$.
Note moreover that the point at which $\gamma_2$ becomes large enough for $\mbox{L2}$ to be preferred depends on which effect we seek to estimate.
When $\gamma_2$ equals 0.15 or 0.16, $\mbox{L1}$ gives a lower MAD for the short-run effect while $\mbox{L2}$ gives a lower MAD for the long-run effect.
Because it is subject to random model selection errors, the GFIC can never outperform the oracle estimator that uses $\mbox{L1}$ when it is optimal in terms of MAD and $\mbox{L2}$ otherwise.
Instead, the GFIC represents a compromise between two extremes: its MAD is never as large as that of the worst specification and never as small as that of the best specification.
When there are large MAD differences between $\mbox{L1}$ and $\mbox{L2}$, however, GFIC is generally close to the optimum.

\subsection{Model and Moment Selection for the Short-run Effect}
\label{sec:Dpanel_sim_SR}
We now consider a more complicated simulation experiment that simultaneously selects over lag specification and endogeneity assumptions. 
In this simulation our target parameter is the short-run effect of $x$ on $y$, as in our empirical example below and the true model contains one lag.
Specifically, 
\[
  y_{it} = \theta x_{it} + \gamma y_{it-1}  + \eta_i + v_{it}
\]
where $i = 1, \dots, n$ and $t = 1, \dots, T$ and the regressor, individual effect and error term are generated according to Equation \ref{eq:covar}.
Our model selection decision example is whether to set $\gamma = 0$ and estimate a specification wihout the lagged dependent variable, while our moment selection decision is whether to use only the instrument set $\mathbf{z}_{it}(\ell,\text{P})$, 
which assumes that $x$ is predetermined, or the instrument set $\mathbf{z}_{it}(\ell,\text{S})$ which assumes that it is strictly exogenous.
Both instrument sets are defined in Equation \ref{eq:Zdpanel}.
We consider four specifications, each estimated by TSLS using the expressions from section \ref{sec:Dpanel}.
The correct specification, $\text{LP}$, estimates both $\gamma$ and $\theta$ using only the ``predetermined'' instrument set.
In contrast, $\text{LS}$ estimates both parameters using the ``strict exogeneity'' instrument set.
The specifications $\text{P}$ and $\text{S}$ set $\gamma=0$ and estimate only $\theta$, using the predetermined and strictly exogenous instrument sets, respectively.
Our simulation design sets $\theta = 0.5$, $\sigma_{x\eta}=0.2$ and varies  $\gamma$, $\sigma_{xv}$, $T$ and $n$ over a grid.
Specifically, we take $\gamma, \sigma_{xv} \in \{0, 0.005, 0.01, \hdots, 0.195, 0.2\}$, $n \in \{250,500\}$, $T \in \{4,5\}$.\footnote{Setting $T$ no smaller than 4 ensures that MSE exists for all four estimators: the finite sample moments of the TSLS estimator only exist up to the order of over-identification.}
All values are computed based on 2000 simulation replications.

Table \ref{tab:Dpanel_RMSE}, presents RMSE values multiplied by 1000 for ease of reading for each of the fixed specifications -- LP, LS, P, and S -- and for the various selection procedures.\footnote{In the interest of space, Table \ref{tab:Dpanel_RMSE} uses a coarser simulation grid than Figure \ref{fig:best}. The supplementary figures in Online Appendix \ref{sec:simulation_supplement} present results over the full simulation grid.} 
We see that there are potentially large gains to be had by intentionally using a mis-specified estimator.
Indeed, the correct specification, $\text{LP}$, is only optimal when both $\rho_{xv}$ and $\gamma$ are fairly large relative to sample size.
When $T=4$ and $n=250$, for example, $\gamma$ and $\rho_{xv}$ must both exceed 0.10 before $\text{LP}$ has the lowest RMSE.
Moreover, the advantage of the mis-specified estimators can be substantial.

\begin{sidewaystable}[htpb]
  \footnotesize
  \centering
  \begin{tabular}{cccc|cccc|cccc|cccc|cccc|cccc} 
 \hline \hline 
\multicolumn{4}{c}{}&\multicolumn{4}{c}{GFIC}&\multicolumn{4}{c}{LP}&\multicolumn{4}{c}{LS}&\multicolumn{4}{c}{P}&\multicolumn{4}{c}{S}\\ 
 \hline
 &  &  & $\rho_{xv}$ & 0 & 0.05 & 0.1 & 0.15 & 0 & 0.05 & 0.1 & 0.15 & 0 & 0.05 & 0.1 & 0.15 & 0 & 0.05 & 0.1 & 0.15 & 0 & 0.05 & 0.1 & 0.15 \\
$T$ & $N$ & $\gamma$ &  &  &  &  &  &  &  &  &  &  &  &  &  &  &  &  &  &  &  &  &  \\
\hline
4 & 250 & 0 &  & 57 & 60 & 62 & 67 & 72 & 70 & 69 & 68 & 51 & 58 & 74 & 95 & 51 & 53 & 52 & 52 & 42 & 49 & 65 & 86 \\
 &  & 0.05 &  & 60 & 62 & 67 & 69 & 74 & 71 & 71 & 70 & 52 & 59 & 77 & 95 & 58 & 57 & 58 & 56 & 43 & 54 & 72 & 91 \\
 &  & 0.1 &  & 66 & 67 & 74 & 75 & 77 & 72 & 74 & 71 & 52 & 57 & 76 & 97 & 76 & 73 & 71 & 70 & 47 & 60 & 77 & 97 \\
 &  & 0.15 &  & 71 & 75 & 78 & 82 & 81 & 77 & 74 & 75 & 53 & 62 & 76 & 99 & 99 & 96 & 92 & 89 & 55 & 69 & 84 & 103 \\
 \hline
 & 500 & 0 &  & 40 & 43 & 48 & 48 & 51 & 49 & 49 & 48 & 37 & 47 & 67 & 90 & 37 & 37 & 36 & 36 & 30 & 40 & 59 & 83 \\
 &  & 0.05 &  & 43 & 47 & 51 & 49 & 52 & 50 & 51 & 49 & 36 & 47 & 67 & 90 & 45 & 45 & 44 & 43 & 31 & 47 & 65 & 87 \\
 &  & 0.1 &  & 45 & 53 & 56 & 54 & 52 & 51 & 50 & 50 & 36 & 47 & 69 & 90 & 67 & 64 & 62 & 59 & 37 & 53 & 73 & 92 \\
 &  & 0.15 &  & 50 & 56 & 59 & 56 & 56 & 53 & 52 & 50 & 36 & 48 & 69 & 92 & 92 & 90 & 85 & 83 & 45 & 63 & 80 & 100 \\
 \hline
5 & 250 & 0 &  & 45 & 48 & 54 & 56 & 56 & 56 & 55 & 54 & 42 & 51 & 70 & 91 & 44 & 45 & 44 & 45 & 36 & 44 & 62 & 83 \\
 &  & 0.05 &  & 48 & 52 & 56 & 55 & 58 & 56 & 55 & 54 & 43 & 52 & 70 & 92 & 52 & 51 & 51 & 48 & 38 & 50 & 68 & 89 \\
 &  & 0.1 &  & 51 & 57 & 61 & 58 & 59 & 57 & 55 & 55 & 44 & 53 & 72 & 94 & 68 & 66 & 65 & 62 & 42 & 57 & 75 & 95 \\
 &  & 0.15 &  & 55 & 61 & 65 & 64 & 60 & 60 & 58 & 57 & 44 & 52 & 74 & 94 & 94 & 89 & 85 & 81 & 51 & 64 & 83 & 100 \\
 \hline
 & 500 & 0 &  & 33 & 36 & 40 & 38 & 41 & 40 & 38 & 38 & 31 & 42 & 63 & 86 & 32 & 31 & 32 & 32 & 27 & 36 & 56 & 79 \\
 &  & 0.05 &  & 35 & 40 & 40 & 38 & 41 & 39 & 38 & 38 & 31 & 42 & 63 & 87 & 42 & 40 & 38 & 38 & 29 & 43 & 62 & 85 \\
 &  & 0.1 &  & 37 & 44 & 44 & 41 & 42 & 40 & 39 & 40 & 31 & 43 & 66 & 88 & 63 & 60 & 58 & 55 & 35 & 52 & 72 & 91 \\
 &  & 0.15 &  & 38 & 45 & 44 & 42 & 42 & 42 & 39 & 40 & 31 & 44 & 67 & 90 & 88 & 85 & 80 & 76 & 44 & 62 & 80 & 98 \\
\hline
\end{tabular}

  \vspace{2em}
  \begin{tabular}{cccc|cccc|cccc|cccc|cccc|cccc} 
 \hline \hline 
\multicolumn{4}{c}{}&\multicolumn{4}{c}{J-test 5\%}&\multicolumn{4}{c}{J-test 10\%}&\multicolumn{4}{c}{GMM-BIC}&\multicolumn{4}{c}{GMM-AIC}&\multicolumn{4}{c}{GMM-HQ}\\ 
 \hline
 &  &  & $\rho_{xv}$ & 0 & 0.05 & 0.1 & 0.15 & 0 & 0.05 & 0.1 & 0.15 & 0 & 0.05 & 0.1 & 0.15 & 0 & 0.05 & 0.1 & 0.15 & 0 & 0.05 & 0.1 & 0.15 \\
$T$ & $N$ & $\gamma$ &  &  &  &  &  &  &  &  &  &  &  &  &  &  &  &  &  &  &  &  &  \\
\hline
4 & 250 & 0 &  & 44 & 52 & 66 & 80 & 46 & 53 & 68 & 77 & 44 & 51 & 68 & 89 & 52 & 58 & 72 & 81 & 46 & 55 & 70 & 88 \\
 &  & 0.05 &  & 47 & 55 & 72 & 88 & 49 & 56 & 73 & 86 & 46 & 55 & 73 & 93 & 53 & 60 & 75 & 88 & 48 & 56 & 74 & 93 \\
 &  & 0.1 &  & 55 & 61 & 78 & 95 & 58 & 63 & 78 & 94 & 50 & 60 & 78 & 98 & 59 & 64 & 79 & 92 & 53 & 60 & 78 & 97 \\
 &  & 0.15 &  & 68 & 74 & 86 & 102 & 73 & 75 & 86 & 101 & 57 & 69 & 83 & 103 & 66 & 73 & 84 & 100 & 59 & 69 & 84 & 102 \\
 \hline
 & 500 & 0 &  & 32 & 41 & 55 & 61 & 34 & 43 & 55 & 57 & 31 & 42 & 61 & 85 & 37 & 47 & 58 & 61 & 34 & 44 & 63 & 79 \\
 &  & 0.05 &  & 34 & 47 & 64 & 74 & 36 & 48 & 63 & 69 & 32 & 47 & 66 & 88 & 39 & 49 & 64 & 67 & 34 & 47 & 66 & 82 \\
 &  & 0.1 &  & 46 & 55 & 71 & 86 & 49 & 56 & 71 & 82 & 39 & 53 & 73 & 92 & 44 & 54 & 70 & 78 & 40 & 53 & 73 & 89 \\
 &  & 0.15 &  & 68 & 67 & 81 & 97 & 73 & 69 & 81 & 95 & 43 & 62 & 80 & 100 & 52 & 61 & 78 & 92 & 43 & 60 & 79 & 98 \\
 \hline
5 & 250 & 0 &  & 37 & 45 & 62 & 74 & 38 & 46 & 61 & 71 & 40 & 49 & 68 & 89 & 43 & 50 & 66 & 75 & 40 & 50 & 69 & 86 \\
 &  & 0.05 &  & 41 & 51 & 67 & 82 & 42 & 52 & 66 & 78 & 42 & 52 & 70 & 92 & 44 & 54 & 68 & 79 & 42 & 52 & 70 & 89 \\
 &  & 0.1 &  & 47 & 58 & 74 & 91 & 50 & 59 & 74 & 88 & 44 & 56 & 74 & 95 & 48 & 58 & 73 & 87 & 45 & 56 & 74 & 94 \\
 &  & 0.15 &  & 63 & 67 & 82 & 97 & 67 & 68 & 82 & 96 & 46 & 58 & 80 & 99 & 54 & 61 & 79 & 93 & 47 & 58 & 80 & 98 \\
 \hline
 & 500 & 0 &  & 28 & 37 & 51 & 51 & 29 & 38 & 49 & 46 & 30 & 41 & 62 & 84 & 31 & 41 & 53 & 50 & 30 & 41 & 60 & 73 \\
 &  & 0.05 &  & 31 & 44 & 59 & 66 & 32 & 44 & 57 & 61 & 31 & 44 & 64 & 86 & 33 & 44 & 57 & 56 & 31 & 44 & 63 & 77 \\
 &  & 0.1 &  & 41 & 52 & 70 & 83 & 44 & 53 & 69 & 78 & 32 & 48 & 71 & 90 & 37 & 49 & 66 & 70 & 33 & 48 & 70 & 86 \\
 &  & 0.15 &  & 65 & 65 & 79 & 94 & 70 & 66 & 78 & 90 & 33 & 52 & 76 & 96 & 41 & 55 & 73 & 84 & 33 & 53 & 76 & 94 \\
\hline
\end{tabular}

  \caption{RMSE values multiplied by 1000 for the simulation experiment from Section \ref{sec:Dpanel_sim_SR}.}
  \label{tab:Dpanel_RMSE}
\end{sidewaystable}

In practice, of course, we do not know the values of $\gamma$, $\rho$, $\theta$, or the other parameters of the DGP so this comparison of finite-sample RMSE values is infeasible.
Instead, we consider using the GFIC to select between the four specifications.
Clearly there are gains to be had from estimating a mis-specified model in certain situations.
The questions remains, can the GFIC identify them?
Because it is an efficient rather than consistent selection criterion, the GFIC remains random, even in the limit.
This means that the GFIC can never outperform the ``oracle'' estimator that uses whichever specification gives the lowest finite-sample RMSE.
Moreover, because our target parameter is a scalar, Stein-type results do not apply: the post-GFIC estimator cannot provide a uniform improvement over the true specification $\text{LP}$.
Nevertheless, the post-GFIC estimator can provide a substantial improvement over $\text{LP}$ when $\rho_{xv}$ and $\gamma$ are relatively small relative to sample size, as shown in the two leftmost panes of the top panel in Table \ref{tab:Dpanel_RMSE}.
This is precisely the situation for which the GFIC is intended: a setting in which we have reason to suspect that mis-specification is fairly mild.
Moreover, in situations where $\text{LP}$ has a substantially lower RMSE than the other estimators, the post GFIC-estimator's performance is comparable.


To provide a basis for comparison, we now consider results for a number of alternative selection procedures. 
The first is a ``Downward J-test,'' which is intended to approximate what applied researchers may do in practice when faced with a model and moment selection decision such as this one.
The Downward J-test selects the \emph{most restrictive} specification that is not rejected by a J-test test with significance level $\alpha \in \left\{ 0.05, 0.1 \right\}$.
We test the specifications $\left\{\text{S}, \text{P}, \text{LS}, \text{LP}\right\}$ \emph{in order} and report the first that is \emph{not rejected}.
This means that we only report $\text{LP}$ if all the other specifications have been rejected.
This procedure is, of course, somewhat \emph{ad hoc} because the significance threshold $\alpha$ is chosen arbitrarily rather than with a view towards some kind of selection optimality. 
We also consider the GMM model and moment selection criteria of \cite{AndrewsLu}.
These take the form
\[
  MMSC_n(b,c) = J_n(b,c) - (|c|-|b|) \kappa_n
\]
where $|b|$ is the number of parameters estimated, $|c|$ the number of moment conditions used, and $\kappa_n$ is a function of $n$. 
Setting $\kappa_n = \log n$ gives the GMM-BIC, while $\kappa_n = 2$ gives the GMM-AIC and $\kappa_n = 2.01 (\log \log n)$ gives the GMM-HQ.
Under certain assumptions, it can be shown that both the GMM-BIC and GMM-HQ are consistent: they select the maximal correctly specified estimator with probability approaching one in the limit. 
To implement these criteria, we calculate the J-test based on the optimal, two-step GMM estimator with a panel robust, heteroscedasticity-consistent, centered covariance matrix estimator for each specification.
To compare selection procedures we use the same simulation grid as above, namely $\gamma, \sigma_{xv} \in \{0, 0.005, 0.01, \hdots, 0.195, 0.20\}$.  
Again, each point on the simulation grid is calculated from 2000 simulation replications. 
The bottom panel of Table \ref{tab:Dpanel_RMSE} presents results for these alternative selection procedures.
There is no clear winner in point-wise RMSE comparisons between the GFIC and its competitors. 
A substantial difference between the GFIC and its competitors emerges, however, when we examine worst-case RMSE.
Here the GFIC clearly dominates, providing the lowest worst-case RMSE across all configurations of $T$ and $n$.
The differences are particularly stark for larger sample sizes.
For example, when $T=5$ and $n=500$ the worst-case RMSE of GFIC is approximately half that of its nearest competitor: GMM-AIC.
The consistent criteria, GMM-BIC and GMM-HQ, perform particularly poorly in terms of worst-case RMSE.
This is unsurprising given that the worst-case risk of any consistent selection criteria diverges as sample size increases.\footnote{See, e.g., \cite{LeebPoetscher2008}.}

\section{Empirical Example: The Demand for Cigarettes}
\label{sec:cigarettes}
We now consider an empirical example illustrating the GFIC in the dynamic panel setting introduced in Section \ref{sec:Dpanel}.
Our exercise is based on \cite{BaltagiEtAl2000} who study the demand for cigarettes using panel data for 46 U.S.\ states between 1963 and 1992. 
Their model is
\[
  \ln C_{it} =  \gamma \ln C_{i,t-1} + \theta \ln P_{it} + \beta_1 \ln Y_{it} + \beta_2 \ln Pn_{it} + \eta_i + \lambda_t + v_{it}
\] 
where $C_{it}$ is the number of packs of cigarettes sold per person aged 16 and above, $P_{it}$ is the real average retail price of a pack of cigarettes, $Y_{it}$ is per capita disposable income, $Pn_{it}$ is the minimum average price of a pack of cigarettes in any state that neighbors state $i$, $\eta_i$ is a state fixed effect, and $\lambda_t$ is a time fixed effect.
The lagged dependent variable in this model is meant to capture habit-persistence in cigarette consumption but it is the price elasticity not the habit-persistence \emph{per se} that is of primary interest. 

\cite{BaltagiEtAl2000} consider an exhaustive list of possible estimators for two target parameters, the short-run price elasticity $\theta$ and the long-run price elasticity $\theta/(1 - \gamma)$, and explore how the resulting estimates vary.
Here we consider selecting between four alternative estimators of the short-run price elasticity $\theta$, as in the second simulation experiment from Section \ref{sec:Dpanel_sim}.
Each specification is estimated by TSLS in first differences, using the expressions from section \ref{sec:Dpanel}.\footnote{Our baseline specification is similar to the estimator that \cite{BaltagiEtAl2000} refer to as FD2SLS. There are two differences however. First, whereas they use lags of the exogenous controls $\ln Y_{it}$ and $\ln Pn_{it}$, our instrument set follows \cite{AndersonHsiao}. 
Second, whereas \cite{BaltagiEtAl2000} appear to have inadvertantly omitted the time dummies from their FD2SLS specification, we include them.}
For simplicity, we assume that the controls $\ln Y_{it}$ and $\ln Pn_{it}$ are exogenous with respect to $v_{it}$.
We focus on two questions.
First: what exogeneity assumption should we impose on $\ln P_{it}$?
Second: should we allow for habit persistence by estimating $\gamma$?

Our baseline specification, $\text{LP}$, estimates both $\gamma$ and $\theta$ and assumes only that $\ln P_{it}$ is predetermined with respect to $v_{it}$. 
This estimator uses the instrument set $\mathbf{z}_{it}(\ell, \text{P})$ with $\ell = 1$ from Equation \ref{eq:Zdpanel}.
For the purposes of this exercise we assume that $\text{LP}$ is correctly specified.
The specification $\text{LS}$ also estimates $\gamma$, but uses the expanded instrument set $\mathbf{z}_{it}(\ell, \text{S})$ with $\ell=1$ from equation \ref{eq:Zdpanel}.
The additional instruments used in $\text{LS}$ are only valid if $\ln P_{it}$ is strictly exogenous with respect to $v_{it}$.
Like $\text{LP}$ and $\text{LS}$, the specifications $\text{P}$ and $\text{S}$ differ in whether or not they impose that $\ln P_{it}$ is predetermined or strictly exogenous.
In contrast, however, they set $\gamma = 0$ and estimate a model with no habit persistence ($\ell =0$).
This increases the number of time periods available for estimation.

\begin{table}[htbp]
  \centering
    \begin{subtable}[h]{0.45\textwidth}
        \centering
     \caption{1975--1980 ($T=6$)}
     \label{tab:cigarettesShort}
     \begin{tabular}{lrrrr}\hline\hline 
         & \multicolumn{1}{c}{$\text{LP}$} & \multicolumn{1}{c}{$\text{LS}$} 
          & \multicolumn{1}{c}{$\text{P}$} & \multicolumn{1}{c}{$\text{S}$}\\
        \hline
        $\widehat{\theta}$ & -0.68 & -0.32 &  -0.28 &  -0.37\\
        Var.\ & 0.16 &  0.02 & 0.07 & 0.01\\ 
        Bias$^2$ & \multicolumn{1}{c}{---} & -4.20 & 0.01 & -3.56 \\
        GFIC  & 0.16 & -4.18 & 0.08  & -3.54  \\
        GFIC+  & 0.16 &  0.02 & 0.08  & 0.01    \\
        \hline
      \end{tabular}
    \end{subtable}
    ~
    \begin{subtable}[h]{0.45\textwidth}
      \centering
     \caption{1975--1985 ($T=11$)}
     \label{tab:cigarettesLong}
     \begin{tabular}{lrrrr}\hline\hline 
         & \multicolumn{1}{c}{$\text{LP}$} & \multicolumn{1}{c}{$\text{LS}$} 
          & \multicolumn{1}{c}{$\text{P}$} & \multicolumn{1}{c}{$\text{S}$}\\
        \hline
        $\widehat{\theta}$ & -0.30 & -0.26 &  -0.38 &  -0.28\\
        Var.\ & 0.06 & 0.01 & 0.05 & 0.01\\ 
        Bias$^2$ & \multicolumn{1}{c}{---} & 2.21 & 0.03 & 1.29\\
        GFIC  &0.06 & 2.22 & 0.08 &  1.30\\
        GFIC+  & 0.06 & 2.22 & 0.08  & 1.30  \\
           \hline
      \end{tabular}
    \end{subtable}
    \caption{Estimates and GFIC values for the price elasticity of demand for cigarettes example from Section \ref{sec:cigarettes} under four alternative specifications. Panel (a) presents results using data from 1975--1980 while Panel (b) presents results using data from 1975--1985. 
GFIC+ gives an alternative version of the GFIC in which a negative squared bias estimate is set equal to zero.}
    \label{tab:cigarettes}
\end{table}

Estimates and GFIC results for all specifications appear in Table \ref{tab:cigarettes}.
In Panel \ref{tab:cigarettesShort} we use data from 1975--1980 only ($T=6$).
After first-differencing, this leaves 4 time periods for estimation in specifications that include a lag ($\text{LP}$ and $\text{LS}$) versus 5 for those that do not ($\text{P}$ and $\text{S}$).
In panel \ref{tab:cigarettesLong} we use data from 1975--1985 ($T=11$).
After first-differencing this leaves 9 time periods for estimation without a lag versus 10 for estimation with a lag.
We choose to artificially shorten the time dimension of the panel from \cite{BaltagiEtAl2000} to illustrate a key feature of the GFIC, namely that it takes into account the number of available time periods when selecting over parameter restrictions and moment conditions.\footnote{Appendix \ref{sec:append_empirical} presents additional results, including the full-sample estimates, and some further discussion.} 
Each column in Table \ref{tab:cigarettes} refers to a particular specification: $\text{LP}$, $\text{LS}$, $\text{P}$ or $\text{S}$.
The first row of each panel gives the associated estimate of the target parameter $\theta$ while the second gives the estimated asymptotic variance of $\sqrt{n}(\widehat{\theta} - \theta_0)$, one of the two ingredients of the GFIC.\footnote{We estimate the asymptotic variance matrix as in \cite{BaltagiEtAl2000}.}
Unsurprisingly the asymptotic variance decreases with the number of time periods available for estimation: for a given sample period $\text{LP}$ and $\text{LS}$ have a higher asymptotic variance than $\text{P}$ and $\text{S}$, and all the estimates based on the 1975--1985 sample are more precise than their counterparts for the 1975-1980 sample.
Moreover, for a given sample period, the estimators with a large instrument set show a lower asymptotic variance: $\text{LS}$ is more precisely estimated than $\text{LP}$ and $\text{S}$ is more precisely estimated than $\text{P}$.

The third row of each panel gives our estimate of the squared asymptotic bias of the various estimators of $\theta$.
The ``---'' entry for the $\text{LP}$ estimator indicates that the GFIC is constructed under the assumption that this specification has no asymptotic bias.
Note that the squared bias estimator is \emph{negative} for $\text{LS}$ and $\text{S}$ in the 1975--1980 sample.
This occurs when the second term in the estimate of the bias matrix $\widehat{B}$ from Corollary \ref{cor:biasestimators}, namely $\widehat{\Psi}\widehat{\Omega}\widehat{\Psi}'$, is larger than the first.
Accordingly, we consider two alternative ways of constructing the GFIC from the asymptotic variance and squared bias estimates.
The first, labelled ``GFIC,'' simply adds squared bias and variance.
The second, labelled ``GFIC+,'' first \emph{truncates} a negative squared bias estimate to zero, and then adds the result to the variance estimate.
For the 1975--1980 sample period we find no evidence of appreciable bias in any of the three ``suspect'' specifications: $\text{LS}$, $\text{P}$, and $\text{LP}$.
In contrast, each of these has a substantially small asymptotic variance than $\text{LP}$, so we would select either $\text{LS}$ or $\text{S}$ depending on whether we prefer to use GFIC or GFIC+.\footnote{When GFIC and GFIC+ disagree, we prefer GFIC+ for the same reason that the positive-part Stein estimator is preferred to the ``plain-vanilla'' Stein estimator.}
In the 1975--1985 sample, the situation changes drastically.
Over this longer time period, the relative advantage of $\text{LS}$, $\text{P}$ and $\text{S}$ over $\text{LP}$ in asymptotic variance decreases substantially and we find evidence of substantial bias in both the $\text{LS}$ and $\text{S}$ specifications.
It appears that over these additional time periods, the assumption that $\ln P_{it}$ is strictly exogenous fails.
Interestingly the difference in GFIC values between $\text{LP}$ and $\text{P}$ in this longer sample is negligible.
As $\text{P}$ has a lower GFIC value than $\text{LP}$ in the 1975--1980 sample, it appears that accounting for habit persistence is relatively unimportant in estimating the short-run price elasticity of cigarette demand.


\section{Conclusion}
\label{sec:conclude}
This paper has introduced the GFIC, a proposal to choose moment conditions and parameter restrictions based on the quality of the estimates they provide. 
The GFIC performs well in simulations for our dynamic panel example. 
While we focus here on applications to panel data, the GFIC can be applied to any GMM problem in which a minimal set of correctly specified moment conditions identifies an unrestricted model. 
A possible extension of this work would be to consider risk functions other than MSE, by analogy to \cite{ClaeskensCroux2006} and \cite{ClaeskensHjort2008}.
Another possibility would be to derive a version of the GFIC for GEL estimators. 
Although first-order equivalent to GMM, GEL estimators often exhibit superior finite-sample properties and may thus improve the quality of the selection criterion \citep{NeweySmith}.

\singlespacing
\small
\bibliographystyle{elsarticle-harv}
\bibliography{GFIC_refs}

\appendix

\section*{Proofs}
\begin{proof}[Proof of Theorem \ref{thm:asymp}]
By a mean-value expansion around $(\theta_0,\gamma_0)$,
	$$\sqrt{n}\left(\widehat{\beta}(b,c) - \beta_0^{(b)}\right) = - K(b,c)\Xi_c \sqrt{n} f_n(\theta_0,\gamma_0) + o_p(1)$$
  and by Assumption \ref{assump:high-level},
$\sqrt{n} f_n(\theta_0, \gamma_0) - \sqrt{n}E\left[f(Z_{ni},\theta_0, \gamma_0) \right]\overset{d}{\rightarrow} (\mathscr{N}_g', \mathscr{N}_h')'$. 
Now, by a mean-value expansion around $\gamma_n$,
	\begin{align*}
		\sqrt{n}E\left[ f(Z_{ni}, \theta_0,\gamma_0) \right] &= \sqrt{n}E\left[ f(Z_{ni}, \theta_0, \gamma_n) \right] + \sqrt{n}\nabla_{\gamma'}E\left[ f(Z_{ni}, \theta_0,\bar{\gamma}) \right] (\gamma_0 -\gamma_n)\\
			&=\left(\left[ \begin{array}{c} 0\\ \tau\end{array}\right] - \nabla_{\gamma'}E\left[ f(Z_{ni}, \theta_0, \bar{\gamma}) \right] \delta\right) \rightarrow \left(\left[ \begin{array}{c} 0\\ \tau\end{array}\right] - F_\gamma\delta\right)
	\end{align*}
  so $\sqrt{n}f_n(\theta_0, \gamma_0) \overset{d}{\rightarrow} \mathscr{N} + (0', \tau')' - F_\gamma \delta$.
  The result follows by the continuous mapping theorem.
\end{proof}

\begin{proof}[Proof of Corollary \ref{cor:valid}]
Since $\Xi_c$ picks out only the components corresponding to $g$,
	$$\Xi_c \left(\left[\begin{array}{c} \mathscr{N}_g\\  \mathscr{N}_h\end{array}\right]+ \left[ \begin{array}{c} 0\\ \tau\end{array}\right] - F_\gamma\delta\right) =  \mathscr{N}_g - G_\gamma \delta.$$
Thus, $\sqrt{n}\left( \widehat{\beta}_v - \beta_0 \right) \overset{d}{\rightarrow} -K_v\left(\mathscr{N}_g - G_\gamma \delta\right)$.
Finally, by the definition of a matrix inverse,
	\begin{align*}
		K_v G_\gamma \delta
    &= \left(\left[\begin{array}{c}G_\theta' \\ G_\gamma'\end{array}\right] W_{gg} \left[\begin{array}{cc} G_\theta & G_\gamma \end{array}\right] \right)^{-1} \left[\begin{array}{c}G_\theta' \\ G_\gamma'\end{array}\right]W_{gg} G_\gamma\delta\\ \\
			&= \left[\begin{array}{cc} 
				G_\theta'W_{gg}G_\theta & G_\theta'W_{gg}G_\gamma\\
				G_\gamma'W_{gg}G_\theta & G_\gamma'W_{gg}G_\gamma
			\end{array} \right]^{-1}
			\left[\begin{array}{c}
				G_\theta'W_{gg}G_\gamma\\
				G_\gamma'W_{gg}G_\gamma
			\end{array}\right]\delta =\left[\begin{array}{c} 0\\ \delta\end{array}\right].
	\end{align*}
\end{proof}

\begin{proof}[Proof of Corollary \ref{cor:mulimit}]
For some $\bar{\gamma}$ between $\gamma_0$ and $\gamma_n = \gamma_0 +\delta/\sqrt{n}$. 
		\[\mu_n  = \varphi(\theta_0, \gamma_0)+ \nabla_\gamma \varphi(\theta_0, \bar{\gamma})'(\gamma_n - \gamma_0) = \mu_0 + \nabla_\gamma \varphi(\theta_0, \bar{\gamma})' \delta/\sqrt{n}\]
by a mean-value expansion.
Hence, $\sqrt{n}(\mu_n - \mu_0) = \nabla_\gamma \varphi(\theta_0, \bar{\gamma})' \delta \rightarrow \nabla_\gamma \varphi(\theta_0, \gamma_0)' \delta$.
The result follows since $\sqrt{n}\left(\widehat{\mu}(b,c) - \mu_n \right)  = \sqrt{n}\left( \widehat{\mu}(b,c) - \mu_0 \right) - \sqrt{n}\left(\mu_n - \mu_0\right)$.
\end{proof}

\begin{proof}[Proof of Corollary \ref{cor:muvalid}]
The result follows from Corollaries \ref{cor:valid} and \ref{cor:mulimit} since,
	\begin{eqnarray*}
		\sqrt{n}\left( \widehat{\mu}_v - \mu_n\right) &\overset{d}{\rightarrow}&\nabla_\beta\varphi_0' \left\{ \left[\begin{array}{c} 0\\ \delta \end{array}\right] -K_v\mathscr{N}_g  \right\}-\nabla_\gamma \varphi_0' \delta\\ 
			&=& -\nabla_\beta\varphi_0' K_v\mathscr{N}_g + \left[\begin{array}{cc} \nabla_\theta\varphi_0'  & \nabla_\gamma\varphi_0' \end{array}\right]\left[\begin{array}{c} 0\\ \delta\end{array}\right]-\nabla_\gamma \varphi_0' \delta = -\nabla_\beta(\gamma_0,\theta_0)' K_v\mathscr{N}_g.
	\end{eqnarray*}
\end{proof}

\begin{proof}[Proof of Corollary \ref{cor:deltahat}]
  The result follows immediately from Corollary \ref{cor:valid}.
\end{proof}

\begin{proof}[Proof of Lemma \ref{lem:tauhatasymp}]
By a mean-value expansion around $\beta_0 = ( \theta_0', \gamma_0')'$,
\[
  \sqrt{n}h_n\left(\widehat{\beta}_v \right) = \sqrt{n}h_n(\beta_0) + H \sqrt{n} \left(\widehat{\beta}_v - \beta_0\right) + o_p(1).
\]
Now, since
\[
  \sqrt{n}f_n(\theta_0, \gamma_0) \overset{d}{\rightarrow} \left[\begin{array}{c} \mathscr{N}_g\\  \mathscr{N}_h\end{array}\right]+ \left[ \begin{array}{c} 0\\ \tau\end{array}\right] - \left[\begin{array}{c}G_\gamma\\ H_\gamma \end{array}\right]\delta
\]
we have $\sqrt{n}h_n(\theta_0, \gamma_0)\overset{d}{\rightarrow} \mathscr{N}_h + \tau - H_\gamma \delta$.
Substituting, 
	\begin{align*}
		\sqrt{n}h_n(\widehat{\beta}_v) &\overset{d}{\rightarrow}  \mathscr{N}_h + \tau - H_\gamma \delta+ H\left( -K_v \mathscr{N}_g + \left[\begin{array}{c} 0 \\ \delta \end{array} \right]\right)
			= \tau - HK_v \mathscr{N}_g + \mathscr{N}_h.
	\end{align*}
\end{proof}

\begin{proof}[Proof of Corollary \ref{cor:biasestimators}]
  Define $(U', V')' = (\delta', \tau')' + \Psi (\mathscr{N}_g', \mathscr{N}_h')'$. 
By the Continuous Mapping Theorem and Theorem \ref{thm:jointbias},
	$$\left[\begin{array}{c} \widehat{\delta} \\ \widehat{\tau} \end{array}\right]\left[\begin{array}{cc} \widehat{\delta}' & \widehat{\tau}'\end{array} \right] \overset{d}{\rightarrow} \left[\begin{array}{c} U\\V \end{array}\right]\left[\begin{array}{cc}U' & V'\end{array} \right] $$
  The result follows since $\Psi\Omega\Psi = E [(U', V')' (U', V')] - (\delta', \tau')' (\delta', \tau')$.
\end{proof}

\begin{proof}[Proof of Theorem \ref{thm:sim}]
To simplify the argument let $\xi = (\delta', \tau')'$ and $\mathscr{M} = \xi + \Psi \mathscr{N}$.
Now, let $a(\xi^*)$ and  $b(\xi^*)$ be quantiles of the distribution of $\Lambda$ calculated under the assumption that the true bias parameter $\xi$ is equal to $\xi^*$ such that $P\{a(\xi^*) \leq \Lambda(\xi^*)\leq b(\xi^*)\} = 1-\alpha_2$.
Define
\[
  a_{min}(\mathscr{M}) = \min\left\{a(\xi^*) \colon \xi^* \in \mathscr{R}(\mathscr{M}|\alpha_1)  \right\}, \quad
  b_{max}(\mathscr{M}) = \max\left\{b(\xi^*) \colon \xi^* \in \mathscr{R}(\mathscr{M}|\alpha_1)  \right\}
\]
and
\[
  \mathscr{R}(\mathscr{M}|\alpha_1) = \left\{ \xi^* \colon \left[\left(\mathscr{M} - \xi^*\right)' \left( \Psi \Omega \Psi' \right)^{-1}\left( \mathscr{M} - \xi^* \right)\right] \leq \chi^2_{p+q}(\alpha_1) \right\}
\]
where $\chi_{p+q}^2(\alpha_1)$ is the $1-\alpha_1$ quantile of a $\chi_{p+q}^2$ distribution. 
By Theorem \ref{thm:jointbias} and Corollary \ref{cor:avg}, 
\[
P\{ \mu_0 \in CI_{sim}\} \rightarrow P\{ a_{min} \leq \Lambda(\delta, \tau) \leq b_{max}\}.
\]
Let $A = \left \{\xi \in \mathscr{R}(\mathscr{M}|\alpha_1) \right\}$.
By construction $P(A) = 1 - \alpha_1$.
Decomposing $P\left\{ a(\xi^*) \leq \Lambda(\xi^*) \leq b(\xi^*)  \right\}$ into the sum of mutually exclusive events, we have
\[
P\left[\left\{ a(\xi^*) \leq \Lambda(\xi^*) \leq b(\xi^*)  \right\} \cap A \right] + P\left[\left\{ a(\xi^*) \leq \Lambda(\xi^*) \leq b(\xi^*)  \right\} \cap A^c \right] = 1-\alpha_2
\]
for every $\xi \in \mathscr{R}(\mathcal{M}|\alpha_1)$.
But since 
$P\left[\left\{ a(\xi^*) \leq \Lambda(\xi^*) \leq b(\xi^*)  \right\} \cap A^c \right] \leq P(A^c) = \alpha_1$,
we see that
$P\left[\left\{ a(\xi^*) \leq \Lambda(\xi^*) \leq b(\xi^*)  \right\} \cap A \right] \geq 1- \alpha_1 - \alpha_2$,
for every $\xi^* \in \mathscr{R}(\mathscr{M}|\alpha_1)$.
By definition, if $A$ occurs then $\xi \in \mathscr{R}(\mathscr{M}|\alpha_1)$ and hence 
$P\left[\left\{ a(\xi) \leq \Lambda(\xi) \leq b(\xi)  \right\} \cap A \right] \geq 1- \alpha_1 - \alpha_2$.
But when $\xi^* \in  \mathscr{R}(\mathscr{M}|\alpha_1)$ we have
$a_{min} \leq a(\xi)$ and $b(\xi) \leq b_{max}$. It follows that 
\[
\left \{ a(\xi) \leq \Lambda (\xi) \leq b(\xi) \right \} \cap A \subseteq \left\{ a_{min} \leq \Lambda (\xi) \leq b_{max}\right \}
\]
and therefore
$1-\alpha_1 - \alpha_2 \leq P\left[\left \{ a(\xi) \leq \Lambda (\xi) \leq b(\xi) \right \} \cap A  \right]
\leq P\left[ a_{min} \leq \Lambda(\xi) \leq b_{max} \right]$.
\end{proof}

\begin{proof}[Proof of Theorem \ref{thm:limitDpanel}.]
  This proof is standard, so we provide only a sketch.
Expanding, 
\begin{align*}
  \sqrt{n}\left[\widehat{\beta}(k,\cdot) - \beta\right] &=(0, \mathbf{0}_{k-m}', \boldsymbol{\delta}')' + \widehat{Q}(k,\cdot)\left[ n^{-1/2}Z'(k,\cdot)\Delta \mathbf{v}\right]\\
  \sqrt{n}\left[\widehat{\beta}(r,\cdot) - \beta_{r}\right] &=  \widehat{Q}(r,\cdot)\left\{  \left[n^{-1}Z'(r,\cdot)( L^{k-m+1}\Delta \mathbf{y}^+ , \ldots,  L^{k}\Delta \mathbf{y}^+)\boldsymbol{\delta}\right] + \left[n^{-1/2}Z'(r,\cdot)\Delta \mathbf{v}^{+}\right] \right\}
\end{align*}
The result follows by applying the Lindeberg-Feller CLT to $n^{-1/2}Z'(k,\cdot)\Delta \mathbf{v}$ and $n^{-1/2}Z'(r,\cdot)\Delta \mathbf{v}^{+}$ and an appropriate law of large numbers to each element of $n^{-1} Z'(r,\cdot)(L^{k-m+1} \Delta \mathbf{y}^{+}, \ldots, L^k \Delta \mathbf{y}^{+})$, where $(\cdot)$ is either $\text{P}$ or $\text{S}$ depending on the instrument set used.
\end{proof}

\begin{proof}[Proof of Theorem \ref{thm:DpanelJoint}]
  Expanding $\widehat{\beta}(k,\text{P})$ from Equation \ref{eq:DpanelEstimators} 
  \begin{align*}
    n^{-1/2}X'\left[ \Delta \mathbf{y} - W(k)\widehat{\beta}(k,\text{P})  \right] 
    &= n^{-1/2}X'\left[ \Delta \mathbf{v}  - W(k)\, \widehat{Q}(k,\text{P}) \, n^{-1} Z'(k,\text{P})\Delta \mathbf{v}\right] \\
    &= \left[
    \begin{array}{cc}
      -n^{-1} X'W(k) \, \widehat{Q}(k,\text{P}) & I_{T-k-1}
    \end{array}
  \right] 
  \left[
  \begin{array}{c}
    n^{-1/2}Z'(k,\text{P}) \Delta \mathbf{v} \\
    n^{-1/2} X' \Delta \mathbf{v}
  \end{array}
\right]
  \end{align*}
  using $\Delta \mathbf{y} = W(k) \beta_n + \Delta \mathbf{v}$. 
  Similarly, expanding $\widehat{\beta}(k,\text{P})$ from Equation \ref{eq:DpanelEstimators},
  \[
    \sqrt{n}\left[ \widehat{\beta}(k,\text{P}) \right] = \left[
    \begin{array}{ccc}
      0 & \mathbf{0}_{k-m}' & \boldsymbol{\delta}'  
    \end{array}
  \right]' + \widehat{Q}(k,\text{P}) n^{-1/2}Z'(k,\text{P}) \Delta \mathbf{v}
  \]
  and since $\widehat{\boldsymbol{\delta}}$ is $\sqrt{n}$ times the last $m$ elements of $\widehat{\beta}(k,\text{P})$ and the last $m$ elements of $\beta$ is zero, we have
  \[
    \widehat{\boldsymbol{\delta}} = \boldsymbol{\delta} + \left[
    \begin{array}{ccc}
    0 & \mathbf{0}'_{k-m} & \boldsymbol{\iota}_m'  
    \end{array}
  \right]' \widehat{Q}(k,\text{P}) n^{-1/2} Z'(k,\text{P})\Delta \mathbf{v}.
  \]
  By a law of large numbers $\widehat{Q}(k,\text{P}) \rightarrow_p Q(k,\text{P})$ and $n^{-1}X'W(k) \rightarrow_p \boldsymbol{\xi}' \otimes \boldsymbol{\iota}_{T-k-1}$, hence 
  \[
    \left[
    \begin{array}{c}
      \widehat{\boldsymbol{\delta}} - \boldsymbol{\delta} \\ \widehat{\tau}
    \end{array}
  \right] = \Psi
  \left[
  \begin{array}{c}
    n^{-1/2}Z'(k,\text{P}) \Delta \mathbf{v} \\
    n^{-1/2} X' \Delta \mathbf{v}
  \end{array}
\right] + o_p(1).
  \]
  The result follows by applying the Lindeberg-Feller CLT jointly to $n^{-1/2}Z'(k,\text{P})\Delta\mathbf{v}$ and $n^{-1/2}X'\Delta \mathbf{v}$ and noting that $\Pi$ maps $[n^{-1/2}Z'(k,\text{S})\Delta \mathbf{v}]$ to $\left[
  \begin{array}{cc}
    \left\{n^{-1/2} Z'(k,P)\Delta \mathbf{v}\right\}' &
    \left\{n^{-1/2} X'\Delta \mathbf{v}\right\}' 
  \end{array}
\right]'.$
\end{proof}

\newpage

\clearpage
\normalsize
\numberwithin{equation}{section}
\numberwithin{table}{section}
\numberwithin{figure}{section}
\begin{center}
  {\Huge Online Appendix\\}
  \vspace{1em}
  {\Large A Generalized Focused Information Criterion for GMM\\}
  \vspace{2em}
  {\large Minsu Chang, Francis J.\ DiTraglia\\ University of Pennsylvania}
\end{center}

\pagenumbering{arabic}
\renewcommand*{\thepage}{Online Appendix - \arabic{page}}
\normalsize
\section{Additional Examples}
\label{sec:additional}
\subsection{Random Effects versus Fixed Effects Example}
\label{sec:REvsFE}
In this section we consider a simple example in which the GFIC is used to choose between and average over alternative assumptions about individual heterogeneity: Random Effects versus Fixed Effects.
For simplicity we consider the homoskedastic case and assume that any strictly exogenous regressors, including a constant term, have been ``projected out'' so we may treat all random variables as mean zero.
To avoid triple subscripts in the notation, we further suppress the dependence of random variables on the cross-section dimension $n$ except within statements of theorems.
Suppose that
\begin{eqnarray}
  y_{it} &=& \beta x_{it}+ v_{it}\\
  v_{it} &=& \alpha_i + \varepsilon_{it}
  \label{eq:REvsFEmodel}
\end{eqnarray}
for $i = 1, \hdots, n$, $t=1, \hdots, T$ where $\varepsilon_{it}$ is iid across $i,t$ with $Var(\varepsilon_{it}) = \sigma^2_{\varepsilon}$ and $\alpha_i$ is iid across $i$ with $Var\left( \alpha_i \right)=\sigma^2_{\alpha}$.
Stacking observations for a given individual over time in the usual way, let $\mathbf{y}_i = (y_{i1}, \hdots, y_{iT})'$ and define $\mathbf{x}_i, \mathbf{v}_i$ and $\boldsymbol{\varepsilon}_i$ analogously.
Our goal in this example is to estimate $\beta$, the effect of $x$ on $y$.
Although $x_{it}$ is uncorrelated with the time-varying portion of the error term, $Cov(x_{it},\varepsilon_{it})=0$, we are unsure whether or not it is correlated with the individual effect $\alpha_i$. 
If we knew for certain that $Cov(x_{it},\alpha_i)=0$, we would prefer to report the ``random effects'' generalized least squares (GLS) estimator given by
\begin{equation}
  \widehat{\beta}_{GLS} = \left(\sum_{i=1}^{n} \mathbf{x}_i' \widehat{\Omega}^{-1} \mathbf{x}_i\right)^{-1}\left(\sum_{i=1}^{n} \mathbf{x}_i'  \widehat{\Omega}^{-1} \mathbf{y}_i   \right)\\
  \label{eq:REdef}
\end{equation}
where $\widehat{\Omega}^{-1}$ is a preliminary consistent estimator of 
\begin{equation}
  \Omega^{-1} = [Var(\mathbf{v}_i)]^{-1} = \frac{1}{\sigma_\epsilon^2} \left[I_T - \frac{\sigma_\alpha^2}{(T\sigma_\alpha^2 + \sigma_\epsilon^2)} \boldsymbol{\iota}_T\boldsymbol{\iota}_T'\right]
  \label{eq:OmegaInvRE}
\end{equation}
and $I_T$ denotes the $T\times 1$ identity matrix and $\boldsymbol{\iota}_T$ a $T$-vector of ones.
This estimator makes efficient use of the variation between and within individuals, resulting in an estimator with a lower variance.
When $Cov(x_{it},\alpha_i)\neq 0$, however, the random effects estimator is biased. 
Although its variance is higher than that of the GLS estimator, the ``fixed effects'' estimator given by 
\begin{equation}
  \widehat{\beta}_{FE} = \left(\sum_{i=1}^{n} \mathbf{x}_i' Q \mathbf{x}_i\bigg)^{-1}\bigg(\sum_{i=1}^{n} \mathbf{x}_i' Q\mathbf{y}_i   \right),
  \label{eq:FEdef}
\end{equation}
where $Q=I_T - \boldsymbol{\iota}_T\boldsymbol{\iota}_T'/T$, remains unbiased even when $x_{it}$ is correlated with $\alpha_i$. 

The conventional wisdom holds that one should use the fixed effects estimator whenever $Cov(x_{it},\alpha_i)\neq 0$.
If the correlation between the regressor of interest and the individual effect is \emph{sufficiently small}, however, the lower variance of the random effects estimator could more than compensate for its bias in a mean-squared error sense.
This is precisely the possibility that we consider here using the GFIC.
In this example, the local mis-specification assumption takes the form
\begin{equation}
  \sum_{t=1}^T E\left[ x_{it}\alpha_i \right] = \frac{\tau}{\sqrt{n}}
  \label{eq:REvsFElocalmisp}
\end{equation}
where $\tau$ is fixed, unknown constant.
In the limit the random effects assumption that $Cov(x_{it},\alpha_i)=0$ holds, since $\tau/\sqrt{n} \rightarrow 0$.
Unless $\tau=0$, however, this assumption \emph{fails} to hold for any finite sample size.
An asymptotically unbiased estimator of $\tau$ for this example is given by
\begin{equation}
  \widehat{\tau}  =( T\widehat{\sigma}_\alpha^2 + \widehat{\sigma}_\epsilon^2) \left[\frac{1}{\sqrt{n}} \sum_{i=1}^n \mathbf{x}_i' \widehat{\Omega}^{-1} (\mathbf{y}_i - \mathbf{x}_i \widehat{\beta}_{FE})\right]
  \label{eq:REvsFEtau}
\end{equation}
leading to the following result, from which we will construct the GFIC for this example.
\begin{thm}[Fixed versus Random Effects Limit Distributions]
\label{thm:REvsFE}
  Let $\left( \mathbf{x}_{ni}, \alpha_{ni}, \boldsymbol{\varepsilon}_{ni} \right)$ be an iid triangular array of random variables such that $Var(\boldsymbol{\varepsilon}_i|\mathbf{x}_{ni},\alpha_{ni})\rightarrow \sigma_{\varepsilon}^2 I_T$, $E[\mathbf{x}_{ni}'Q\boldsymbol{\varepsilon}_{ni}]=0$, and $E\left[ \alpha_i \boldsymbol{\iota}_T'\mathbf{x}_{ni} \right]=\tau/\sqrt{n}$ for all $n$.
  Then, under standard regularity conditions,
\[
  \left[\begin{array}{c}
\sqrt{n} (\widehat{\beta}_{RE} - \beta)\\
\sqrt{n} (\widehat{\beta}_{FE} - \beta)\\
\widehat{\tau}
\end{array}\right] \overset{d}{\rightarrow}  \left( 
\left[\begin{array}{c}
c\tau \\
0  \\
\tau\\
\end{array}\right],  
\left[\begin{array}{ccc}
\eta^2 & \eta^2 & 0 \\
\eta^2 & c^2\sigma^2 + \eta^2 & -c\sigma^2\\ 
0 & -c\sigma^2 & \sigma^2
\end{array}\right] \right)
\]
where $\eta^2 = E[\mathbf{x}_i'\Omega^{-1}\mathbf{x}_i]$, $c = E[\mathbf{x}_i' Q \mathbf{x}_i]/(T\sigma_\alpha^2 + \sigma_\varepsilon^2)$, and
\[\sigma^2 = \frac{(T\sigma_{\alpha}^2 + \sigma_{\varepsilon}^2)^2}{E\left[ \mathbf{x}_i'\Omega^{-1}\mathbf{x}_i \right]}\left( \frac{\sigma_{\varepsilon}^2}{E\left[ \mathbf{x}_i \Omega^{-1} \mathbf{x}_i \right]E\left[ \mathbf{x}_i Q \mathbf{x}_i \right]} - 1 \right). \]
\end{thm}

\begin{proof}[Proof of Theorem \ref{thm:REvsFE}]
  This proof is standard so we provide only a sketch.
  First, let 
  $A_n = (n^{-1}\sum_{i=1}^n \mathbf{x}_i' \widehat{\Omega}^{-1}\mathbf{x}_i)$, $B_n = (n^{-1} \sum_{i=1}^n \mathbf{x}_i' Q\mathbf{x}_i)$, and $C_n = T\widehat{\sigma}_\alpha^2 + \widehat{\sigma}_\varepsilon^2$.
  Now, expanding $\widehat{\beta}_{FE}$, $\mathbf{\beta}_{RE}$, and $\widehat{\tau}$ and re-arranging
\[
  \left[
  \begin{array}{c}
\sqrt{n} (\widehat{\beta}_{RE} - \beta)\\
\sqrt{n} (\widehat{\beta}_{FE} - \beta)\\
\widehat{\tau}
  \end{array}
\right] = 
\left[
\begin{array}{cc}
  A_n^{-1} & 0 \\
  0 & B_n^{-1} \\
  C_n & -C_nA_nB_n^{-1}
\end{array}
\right] \left[
\begin{array}{c}
  n^{-1/2} \sum_{i=1}^n \mathbf{x}_i' \widehat{\Omega}^{-1} \mathbf{v}_i\\ 
  n^{-1/2} \sum_{i=1}^n \mathbf{x}_i' Q\mathbf{v}_i
\end{array}
\right].
\]
The result follows by applying a law of large numbers to $A_n, B_n, C_n$, and $\widehat{\Omega}$ and the Lindeberg-Feller CLT jointly to $n^{-1/2} \sum_{i=1}^n \mathbf{x}_i'Q\mathbf{v}_i$ and $n^{-1/2}\sum_{i=1}^n \mathbf{x}_i' \Omega^{-1} \mathbf{v}_i$. 
\end{proof}

We see from Theorem \ref{thm:REvsFE} that $AMSE(\widehat{\beta}_{RE}) = c^2 \tau^2 + \eta^2$, $AMSE(\widehat{\beta}_{FE}) = c^2\sigma^2 + \eta^2$, and $\widehat{\tau}^2 - \sigma^2$ provides an asymptotically unbiased estimator of $\tau^2$.
Thus, substituting $\widehat{\tau}^2 - \sigma^2$ for $\tau$ and rearranging the preceding AMSE expressions, the GFIC tells us that we should select the random effects estimator whenever $|\widehat{\tau}|\leq \sqrt{2} \sigma$.
To implement this rule in practice, we construct a consistent estimator of $\sigma^2$, for which we require estimators of $\sigma_{\alpha}^2, \sigma_{\varepsilon}^2$ and $\sigma_{v}^2 = Var(\alpha_i + \varepsilon_{it})$.
We estimate these from the residuals
\[
\widehat{\epsilon}_{it} = (y_{it} -\bar{y}_i) - (x_{it} - \bar{x}_i) \widehat{\beta}_{FE}; \quad
\widehat{v}_{it} = y_{it} - x_{it} \widehat{\beta}_{OLS}
\]
where $\widehat{\beta}_{OLS}$ denotes the \emph{pooled} OLS estimator of $\beta$, leading to the variance estimators 
\[
\widehat{\sigma}_\alpha^2 = \widehat{\sigma}_v^2 - \widehat{\sigma}_\epsilon^2; \quad
\widehat{\sigma}_\epsilon^2 = \frac{1}{n(T-1)-1} \sum_{i=1}^n \sum_{t=1}^T \widehat{\epsilon}_{it}^2; \quad
\widehat{\sigma}_v^2 = \frac{1}{nT-1} \sum_{i=1}^n \sum_{t=1}^T \widehat{v}_{it}^2
\]

Selection, of course, is a somewhat crude procedure: it is essentially an average that uses all-or-nothing weights.
As a consequence, relatively small changes to the data could produce discontinuous changes in the weights, leading to a procedure with a high variance.
Rather than selecting between the random effects and fixed effects estimators based on estimated AMSE, an alternative idea is to consider a more general weighted average of the form
\[\widetilde{\beta}(\omega) =  \omega \widehat{\beta}_{FE} + (1 - \omega)\widehat{\beta}_{RE}\]
and for $\omega \in [0,1]$ \emph{optimize} the choice of $\omega$ to minimize AMSE. 
From Theorem \ref{thm:REvsFE} we see that the AMSE-minimizing value of $\omega$ is $\omega^* = (1 + \tau^2/\sigma^2)^{-1}$.
Substituting our asymptotically unbiased estimator of $\tau^2$ and our consistent estimator $\widehat{\sigma}^2$ of $\sigma^2$, we propose the following plug-in estimator of $\omega^*$ 
\begin{equation*}
  \omega^* = \left[ 1 + \frac{ \max\left\{  \widehat{\tau}^2 - \widehat{\sigma}^2, 0\right\}}{\widehat{\sigma}^2} \right]^{-1}
\end{equation*}
where we take the maximum over $\widehat{\tau}^2 - \widehat{\sigma}^2$ and zero so that $\widehat{\omega}^*$ is between zero and one. 
This proposal is related to the Frequentist Model Average estimators of \cite{hjort2003frequentist} as well as \cite{HansenShrink}, and \cite{DiTraglia2016}.

\subsection{Slope Heterogeneity Example}
\label{sec:slopeHet}
Suppose we wish to estimate the average effect $\beta$ of a regressor $x$ in a panel setting where this effect may vary by individual: say $\beta_i \sim \mbox{iid}$ over $i$.
One idea is to simply ignore the heterogeneity and fit a pooled model.
A pooled estimator will generally be quite precise, but depending on the nature and extent of heterogeneity could show a serious bias.
Another idea is to apply the \emph{mean group estimator} by running separate time-series regressions for each individual and averaging the result over the cross-section \citep{Swamy1970,PesaranSmith1995,PesaranEtAl1999}.
This approach is robust to heterogeneity but may yield an imprecise estimator, particularly in panels with a short time dimension.
To see how the GFIC navigates this tradeoff, consider the following DGP:
	\begin{align}
				y_{it} &= \beta_i x_{it} + \epsilon_{it}\\
        \beta_i &= \beta + \eta_i, \quad \eta_i \sim \mbox{iid} (0, \sigma_\eta^2)
	\end{align}
  where $x_{it}$ is uncorrelated with $\varepsilon_{it}$ but is \emph{not} assumed to be independent of $\eta_i$. 
As in the preceding examples $i = 1, \hdots, n$ indexes individuals, $t=1, \hdots, T$ indexes time periods, and we assume without loss of generality that all random variables are mean zero and any exogenous controls have been projected out.
For the purposes of this example, assume further that $\epsilon_{it}$ is iid over both $i$ and $t$ with variance $\sigma_\epsilon^2$ and that both error terms are homoskedastic: $E[\epsilon_{it}^2 | x_{it}] = \sigma_\epsilon^2$ and $E[\eta_i^2 | x_{it}] = \sigma_\eta^2$, and $E[\eta_i \epsilon_{it} | x_{it}] = 0$.
Neither homoskedasticity nor time-independent errors are required to apply the GFIC to this example, but these assumptions simplify the exposition.  
We place no assumptions on the joint distribution of $x_{it}$ and $\eta_i$.

Stacking observations, let $\mathbf{y}_i = (y_{i1}, \ldots, y_{iT})'$ and define $\mathbf{x}_i$ analogously. 
We consider two estimators: the pooled OLS estimator $\widehat{\beta}_{OLS}$ and the mean-group estimator $\widehat{\beta}_{MG}$
\begin{align}
\widehat{\beta}_{OLS} &= \left(\sum_{i=1}^{n} \mathbf{x}_i'  \mathbf{x}_i \right)^{-1}\left(\sum_{i=1}^{n} \mathbf{x}_i' \mathbf{y}_i   \right)\\ 
\widehat{\beta}_{MG}  &= \frac{1}{n}\sum_{i=1}^n \widehat{\beta}_i 
= \frac{1}{n} \sum_{i=1}^n \left( \mathbf{x}_i'  \mathbf{x}_i\right)^{-1}\left( \mathbf{x}_i'  \mathbf{y}_i   \right)
\end{align}
where $\widehat{\beta}_i$ denotes the OLS estimator calculated using observations for individual $i$ only. 
If we knew with certainty that there was no slope heterogeneity, we could clearly prefer $\widehat{\beta}_{OLS}$ as it is both unbiased and has the lower variance.
In the presence of heterogeneity, however, the situation is more complicated.
If $E[\mathbf{x}_i' \mathbf{x}_i \eta_i]\neq 0$ then $\widehat{\beta}_{OLS}$ will show a bias whereas $\widehat{\beta}_{MG}$ will not.
To encode this idea within the local mis-specification framework, we take $E[\mathbf{x}_i'\mathbf{x}_i \eta_i] = \tau/\sqrt{n}$ so that, for any fixed $n$ the OLS estimator is biased unless $\tau = 0$ but this bias disappears in the limit.
Turning our attention from bias to variance, we might expect that $\widehat{\beta}_{OLS}$ would remain the more precise estimator in the presence of heterogeneity.
In fact, however, this need not be the case: as we show below, $\widehat{\beta}_{MG}$ will have a \emph{lower} variance than $\widehat{\beta}_{OLS}$ if $\sigma_{\eta}^2$ is sufficiently large.
To construct the GFIC for this example, we estimate the bias parameter $\tau$ by substituting the mean group estimator into the OLS moment condition:
\begin{equation}
\widehat{\tau} = \frac{1}{\sqrt{n}} \sum_{i=1}^n \mathbf{x}_i' (\mathbf{y}_i - \mathbf{x}_i \widehat{\beta}_{MG}).
\end{equation}
The key result needed to apply the GFIC in this this example gives the joint limiting distribution of $\widehat{\tau}$, the mean-group estimator, and the OLS estimator.
\begin{thm}[Limit Distribution of OLS and Mean-Group Estimators]
\label{thm:OLSvsMG}
Let $\left( \mathbf{x}_{ni},\eta_{ni}, \boldsymbol{\varepsilon}_{ni} \right)$ be an iid triangular array of random variables such that $E[\mathbf{x}_{ni}'\boldsymbol{\varepsilon}_{ni}] =0$, $Var(\boldsymbol{\varepsilon}_{ni}|\mathbf{x}_{ni}) \rightarrow \sigma_{\varepsilon}^2 I_T$, $Var(\eta_{ni}|\mathbf{x}_{ni}) \rightarrow \sigma_{\eta}^2$, $E[\eta_{ni} \boldsymbol{\varepsilon}_{ni}|\mathbf{x}_{ni}] \rightarrow 0$, and $E[\mathbf{x}_{ni}'\mathbf{x}_{ni}\eta_{ni}] = \tau/\sqrt{n}$.
Then, under standard regularity conditions, 
\[
  \left[\begin{array}{c}
\sqrt{n} (\widehat{\beta}_{OLS} - \beta)\\
\sqrt{n} (\widehat{\beta}_{MG} - \beta)\\
\widehat{\tau}
\end{array}\right] \overset{d}{\rightarrow} N \left( 
\left[\begin{array}{c}
\tau/\kappa \\
0  \\
\tau\\
\end{array}\right],  
\left[
\begin{array}{ccc}
  \displaystyle\left(\frac{\lambda^2 + \kappa^2}{\kappa^2}\right)\sigma_\eta^2 + \frac{\sigma_\varepsilon^2}{\kappa} & \displaystyle \sigma_\eta^2 + \frac{\sigma_\varepsilon^2}{\kappa} &\displaystyle\left(\frac{\lambda^2}{\kappa}\right)\sigma_\eta^2  \\
  & \sigma_{\eta}^2 + \zeta \sigma_\varepsilon^2 & \sigma_{\varepsilon}^2 (1 - \kappa\zeta) \\
   &  & \lambda^2 \sigma_\eta^2 + \kappa(\kappa\zeta - 1) \sigma_\varepsilon^2 
\end{array}
\right]\right)
\]
where $\kappa = E[\mathbf{x}_i' \mathbf{x}_i]$, $\lambda^2 = Var\left( \mathbf{x}_i'\mathbf{x}_i \right)$, and $\zeta = E\left[ \left( \mathbf{x}_i' \mathbf{x}_i \right)^{-1} \right]$.  
\end{thm}

\begin{proof}[Proof of Theorem \ref{thm:OLSvsMG}]
Expanding the definitions of the OLS and mean-group estimators, 
\begin{align*}
  \sqrt{n} (\widehat{\beta}_{OLS} - \beta) &=  \left[\begin{array}{cc}
    \left(n^{-1}\sum_{i=1}^{n} \mathbf{x}_i'  \mathbf{x}_i\right)^{-1} & \left(n^{-1}\sum_{i=1}^{n} \mathbf{x}_i'  \mathbf{x}_i\right)^{-1}\end{array} \right] 
  \left[\begin{array}{c} 
n^{-1/2}\sum_{i=1}^{n} \mathbf{x}_i'\mathbf{x}_i\mathbf{\eta}_i   \\
n^{-1/2}\sum_{i=1}^{n} \mathbf{x}_i'\mathbf{\varepsilon}_i   
\end{array}\right]\\
\sqrt{n} (\widehat{\beta}_{MG} - \beta)  &=  n^{-1/2} \sum_{i=1}^n \left[\eta_i + (\mathbf{x}_i'\mathbf{x}_i)^{-1} \mathbf{x}_i'\varepsilon_i\right]
\end{align*}
and proceeding similarly for $\widehat{\tau}$,
\[
\widehat{\tau}  =  
\left[
  \begin{array}{ccc}
  1 & 1& -n^{-1}\sum_{i=1}^n \mathbf{x}_i'\mathbf{x}_i
\end{array}\right] 
\left[\begin{array}{c}
n^{-1/2} \sum_{i=1}^n \mathbf{x}_i'\mathbf{x}_i \eta_i \\
n^{-1/2} \sum_{i=1}^n \mathbf{x}_i'\varepsilon_i\\
n^{-1/2} \sum_{i=1}^n \left\{ \eta_i + (\mathbf{x}_i'\mathbf{x}_i)^{-1} \mathbf{x}_i'\varepsilon_i\right\} 
\end{array}
\right].
\]
The result follows, after some algebra, by a LLN and the Lindeberg-Feller CLT.
\end{proof}

As mentioned above, the OLS estimator need not have a lower variance than the mean-group estimator if $\sigma_{\eta}^2$ is sufficiently large.
This fact follows as a corollary of Theorem \ref{thm:OLSvsMG}.

\begin{cor}
  Under the conditions of Theorem \ref{thm:OLSvsMG}, the asymptotic variance of the OLS estimator is lower than that of the mean-group estimator if and only if $\lambda^2\sigma_\eta^2 < \sigma_\varepsilon^2(\kappa^2\zeta - \kappa)$, 
where $\kappa = E[\mathbf{x}_i' \mathbf{x}_i]$, $\lambda^2 = Var\left( \mathbf{x}_i'\mathbf{x}_i \right)$, and $\zeta = E\left[ \left( \mathbf{x}_i' \mathbf{x}_i \right)^{-1} \right]$.  
\label{cor:MG}
\end{cor}

Note, as a special case of the preceding, that the OLS estimator is guaranteed to have the lower asymptotic variance when $\sigma_{\eta}^2 = 0$ since $E[\mathbf{x}_i'\mathbf{x}]^{-1} < E[(\mathbf{x}_i'\mathbf{x}_i)^{-1}]$ by Jensen's inequality.
When $\sigma_{\eta}^2 \neq 0$, the situation is in general much more complicated.
A simple normal example, however, provides some helpful intuition.
Suppose that for a given individual the observations $x_{it}$ are iid standard normal over $t$.
Then $\mathbf{x}_i'\mathbf{x}_i \sim \chi^2_T$, so that $\kappa = T$, $\lambda^2 = 2T$ and $\zeta = 1 / (T-2)$, provided of course that $T>2$.
Substituting these into Corollary \ref{cor:MG}, the OLS estimator will have the lower asymptotic variance whenever $(T-2)\sigma^2_\eta < \sigma_{\varepsilon}^2$.
All else equal, the shorter the panel, the more likely that OLS will have the lower variance.
But if $\sigma_{\eta}^2$ is large enough, the length of the panel becomes irrelevant: with enough slope heterogeneity, the mean-group estimator has the advantage both in bias and variance.

To apply the GFIC in practice, we first need to determine whether the OLS estimator has the smaller asymptotic variance.
This requires us to estimate the quantities $\lambda^2, \kappa$, and $\zeta$ from Theorem \ref{thm:OLSvsMG} along with $\sigma_{\eta}^2$ and $\sigma_{\varepsilon}^2$.
The following estimators are consistent under the assumptions of Theorem \ref{thm:OLSvsMG}:
\begin{align*}
  \widehat{\kappa} &= \frac{1}{n}\sum_{i=1}^n \mathbf{x}_i'\mathbf{x}_i &
  \widehat{\zeta} &= \frac{1}{n}\sum_{i=1}^n (\mathbf{x}_i'\mathbf{x}_i)^{-1} \\
  \widehat{\lambda^2} &= \frac{1}{n-1}\sum_{i=1}^n (\mathbf{x}_i'\mathbf{x}_i - \widehat{\kappa})^2  &
\widehat{\sigma}_\epsilon^2 &= \frac{1}{nT - 1} \sum_{i=1}^n \sum_{t=1}^T (y_{it}-x_{it}\widehat{\beta}_{OLS})^2\\
\widehat{\sigma}_\eta^2 &= \frac{S_b}{n-1} -\frac{1}{n} \sum_{i=1}^n \widehat{\sigma}_\epsilon^2 (\mathbf{x}_i'\mathbf{x}_i)^{-1} & 
S_b &= \sum_{i=1}^n \widehat{\beta}_i^2 - n\left(\frac{1}{n} \sum_{i=1}^n \widehat{\beta}_i\right)^2 
\end{align*}
If the estimated asymptotic variance of the mean-group estimator is lower than that of the OLS estimator, then there is no need to estimate AMSE: we should simply use the mean-group estimator.
If this is not the case, then we construct the GFIC using the asymptotically unbiased estimator $\widehat{\tau}^2 - \widehat{\sigma}_\tau^2$ of $\tau^2$, where $\widehat{\sigma}_\tau^2 = \widehat{\lambda}^2 \widehat{\sigma}^2_{\eta} + \widehat{\kappa}(\widehat{\kappa}\widehat{\zeta} - 1) \widehat{\sigma}_{\varepsilon}^2$ is a consistent estimator of the asymptotic variance of $\widehat{\tau}$.

\section{Supplementary Simulation Results}
\label{sec:simulation_supplement}
\subsection{Fixed vs.\ Random Effects Example}
\label{sec:REvsFEsim}
\begin{figure}[htbp]
  \centering
  \input{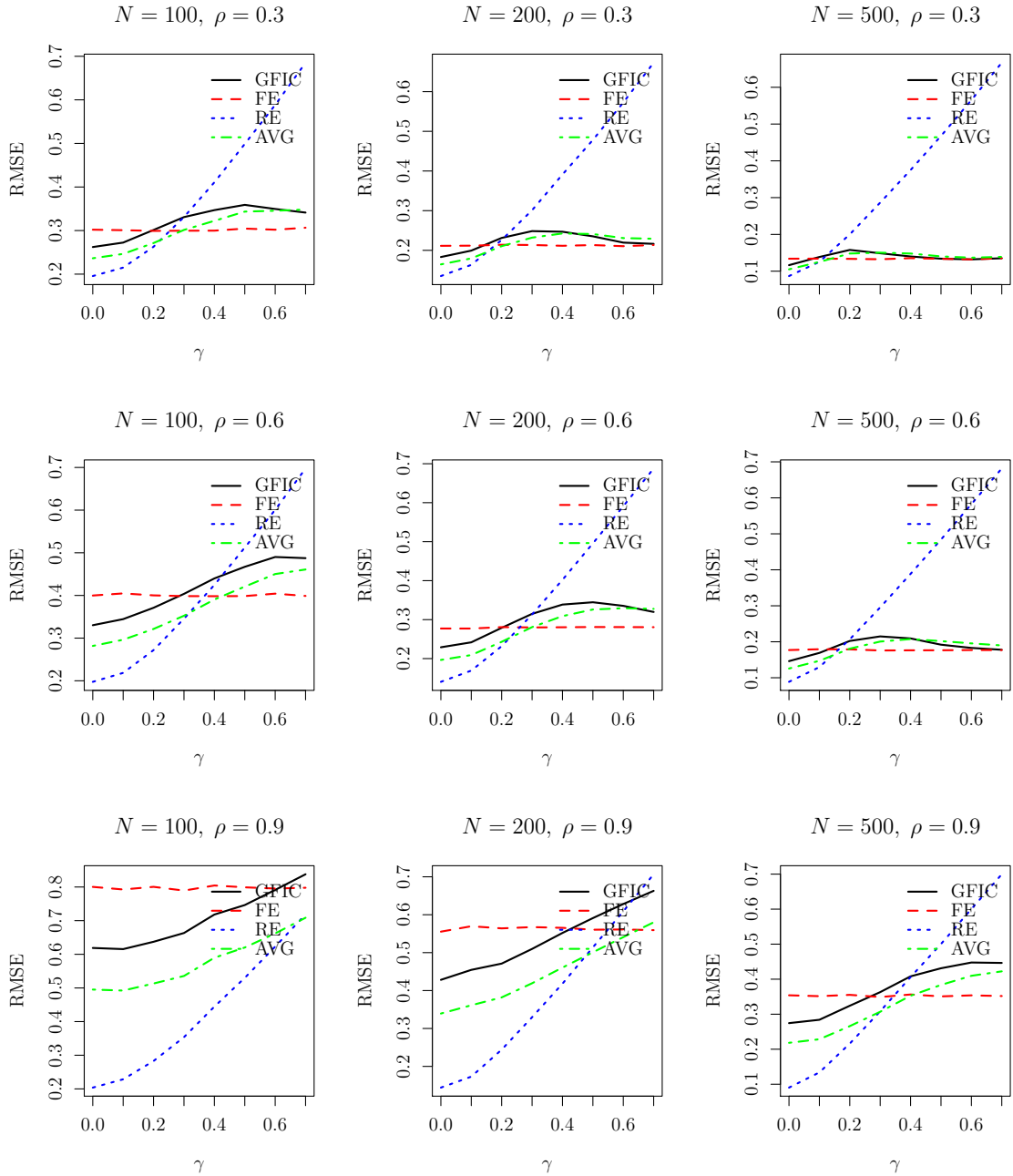}
  \caption{RMSE values for the Random vs.\ Fixed effects simulation example from Section \ref{sec:REvsFEsim} with $T=2$ and  $\sigma_{\varepsilon}^2 = 2.5$. Results are based on 10,000 simulation replications.}
  \label{fig:REvsFE_T2}
\end{figure}

We employ a simulation design similar to that used by \cite{GuggenbergerRE}, namely
\begin{equation*}
  y_{it} = 0.5 x_{it} + \alpha_i + \varepsilon_{it}
\end{equation*}
where
\[
  \begin{bmatrix}
x_{i1}\\
x_{i2}\\
\vdots\\
x_{iT}\\
\alpha_i
\end{bmatrix}  \overset{\mbox{iid}}{\sim} N \left (
\begin{bmatrix}
0\\
0\\
\vdots\\
0\\
0
\end{bmatrix} , \, 
\begin{bmatrix}
1 & \rho & \ldots & \rho & \gamma\\
\rho & 1 & \ldots & \rho & \gamma\\
\vdots & \vdots&  \ddots & \vdots &\vdots\\
\rho & \rho & \ldots & 1 & \gamma\\
\gamma& \gamma &\ldots& \gamma & 1 
\end{bmatrix}
\right)
\]
independently of $\left( \varepsilon_{i1}, \dots, \varepsilon_{iT} \right)' \sim \mbox{ iid } N(0, \sigma_{\varepsilon}^2 \mathbf{I}_T)$.
In this design, $\gamma$ controls the correlation between $x_{it}$ and the individual effects $\alpha_i$, while $\rho$ controls the persistence of $x_{it}$ over time.
Larger values of $\gamma$ correspond to larger violations of the assumption underlying the random effects estimator, increasing its bias.
Larger values of $\rho$, on the other hand, decrease the amount of variation within individuals, thus \emph{increasing} the variance of the fixed effects estimator. 
Figures \ref{fig:REvsFE_T2} presents RMSE values for the random effects GLS estimator and fixed effects estimator along with those for the post-GFIC and averaging estimators described above in Section \ref{sec:REvsFE} over a grid of values for $\gamma$, $\rho$ and $n$ with $T=2$.
Results for $T=5$ appear in Figure \ref{fig:REvsFE_T5} of Appendix \ref{sec:simulation_supplement}.
All calculations are based on 10,000 simulation replications.
In the interest of space, we present only results for $\sigma_{\varepsilon}^2 = 2.5$ and a coarse parameter grid for $\rho$ here. 
Additional results are available upon request.

We see from Figure \ref{fig:REvsFE_T2} that, regardless of the configuration of the other parameter values, there is always a range of values for $\gamma$ for which the random effects estimator has a smaller RMSE than the fixed effects estimator.
The width of this range increases as either the number of individuals $N$ or the number of time periods $T$ decrease.
It also increases as the persistence $\rho$ of $x_{it}$ increases.
Indeed, when $N$ and $T$ are relatively small and $\rho$ is relatively large, the individual effects $\alpha_{i}$ can be \emph{strongly} correlated with $x_{it}$ and still result in a random effects estimator with a lower RMSE than the fixed effects estimator. 
The post-GFIC estimator essentially ``splits the difference'' between the random and fixed effects estimators.
While it cannot provide a uniform improvement over the fixed effects estimator, the post-GFIC estimator performs well.
When $\gamma$ is not too large it can yield a substantially lower RMSE than the fixed effects estimator.
The gains are particularly substantial when $x_{it}$ is relatively persistent and $T$ relatively small, as is common in micro-panel datasets.
The averaging estimator performs even better, providing a nearly uniform improvement over the post-GFIC estimator.
Only at very large values of $\gamma$ does it yield a higher RMSE, and these are points in the parameter space where the fixed effects, post-GFIC and averaging estimators are for all intents and purposes identical in RMSE.
Results for $T=5$ are qualitatively similar.
See Figure \ref{fig:REvsFE_T5} of Appendix \ref{sec:simulation_supplement} for details.
Note that in when $T=5$, setting $\rho = 0.3$ violates positive definiteness so we take $\rho=0.4$ as our smallest value in this case.

The results we have discussed here focus on the comparison of the GFIC to the fixed effects, random effects, and averaging estimators.
One might also wonder how the GFIC compares to a Durbin-Hausman-Wu (DHW) pre-test estimator that reports the random effects estimator unless the difference between $\widehat{\beta}_{FE}$ and $\widetilde{\beta}_{RE}$ is sufficiently large.
By an argument similar to that of \cite{DiTraglia2016} Section 3.2, the GFIC in this particular example is essentially equivalent to a DHW pre-test estimator based on a \emph{particular} significance level dictated by our desire to minimize an asymptotically unbiased estimator of AMSE.
As such, a comparison of the GFIC to a DHW pre-test estimator based on the more standard significance levels $0.1$ and $0.05$ will be qualitatively similar to \cite{DiTraglia2016} Figure 2.
In particular, there is no choice of significance level for which one DHW-based pre-test estimator, including in this case the GFIC, uniformly dominates any other.

\begin{figure}[H]
  \centering
  \input{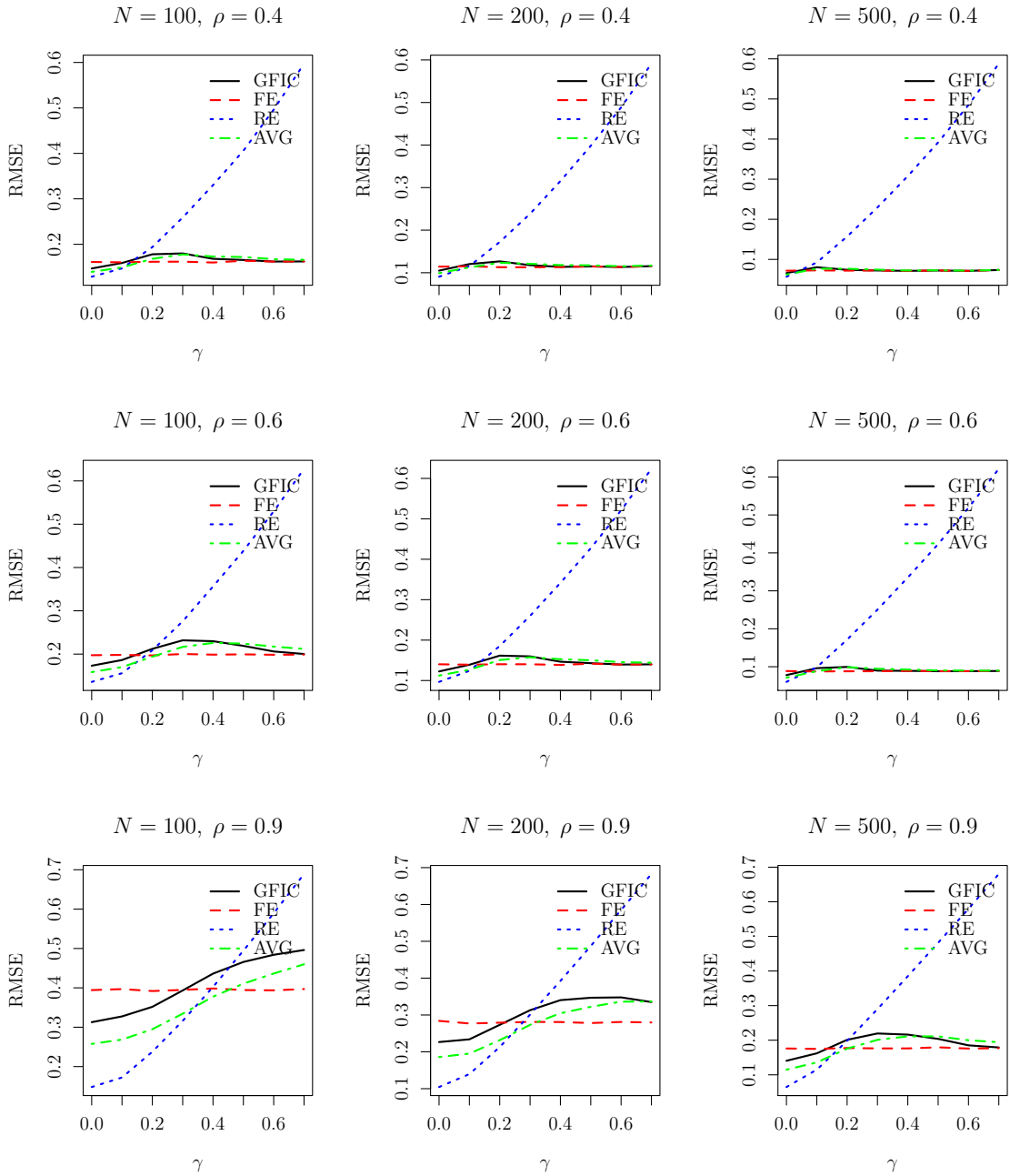}
  \caption{RMSE for Random vs.\ Fixed effects example from Section \ref{sec:REvsFEsim}: $T=5, \sigma_{\varepsilon}^2 = 2.5$}
  \label{fig:REvsFE_T5}
\end{figure}

\subsection{Figures for Dynamic Panel Simulation}
\begin{figure}[htbp]
\centering
\includegraphics[scale = 0.8]{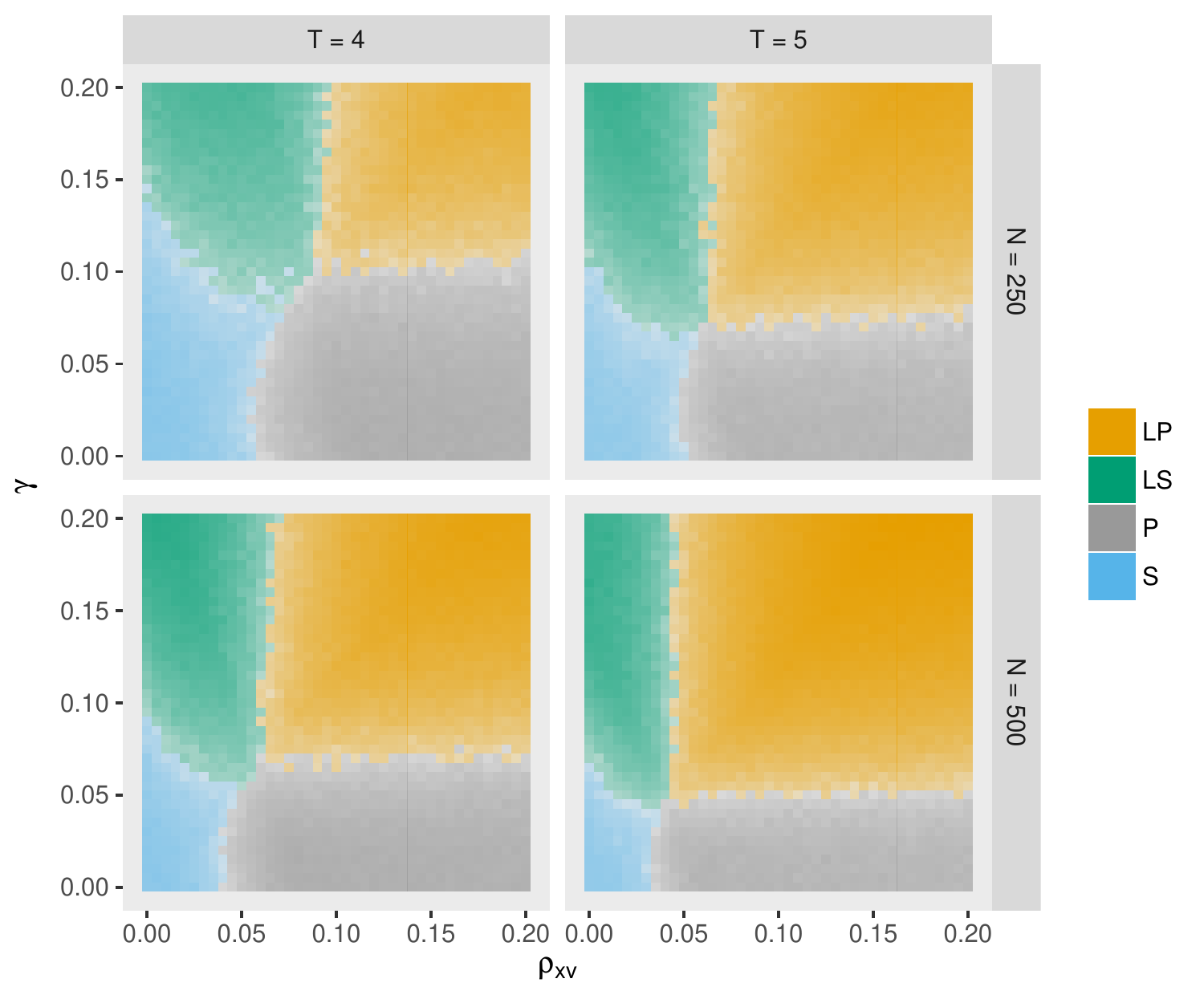}
\caption{Minimum RMSE specification at each combination of parameter values for the simulation experiment from Section \ref{sec:Dpanel_sim_SR}. Color saturation at a given grid point indicates RMSE relative to second best specification.}
\label{fig:best}
\end{figure}

\begin{figure}
\centering
\includegraphics[scale = 0.8]{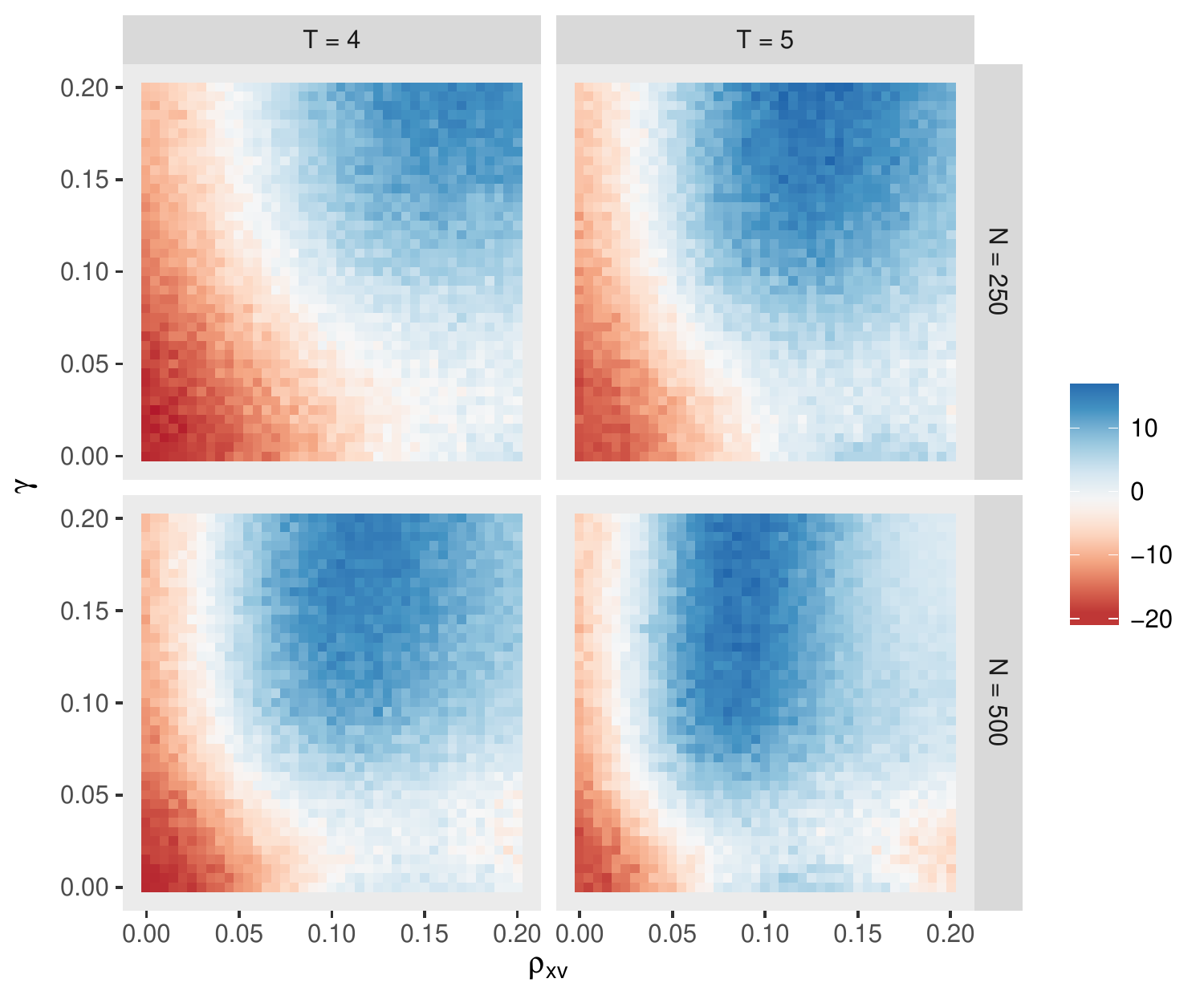}
\caption{RMSE of the post-GFIC estimator relative to that of the true specification ($\text{LP}$) in the dynamic panel simulation experiment from Section \ref{sec:Dpanel_sim_SR}.}
\label{fig:GFIC_rel_LP}
\end{figure}

\begin{figure}
\centering
\includegraphics[scale = 0.8]{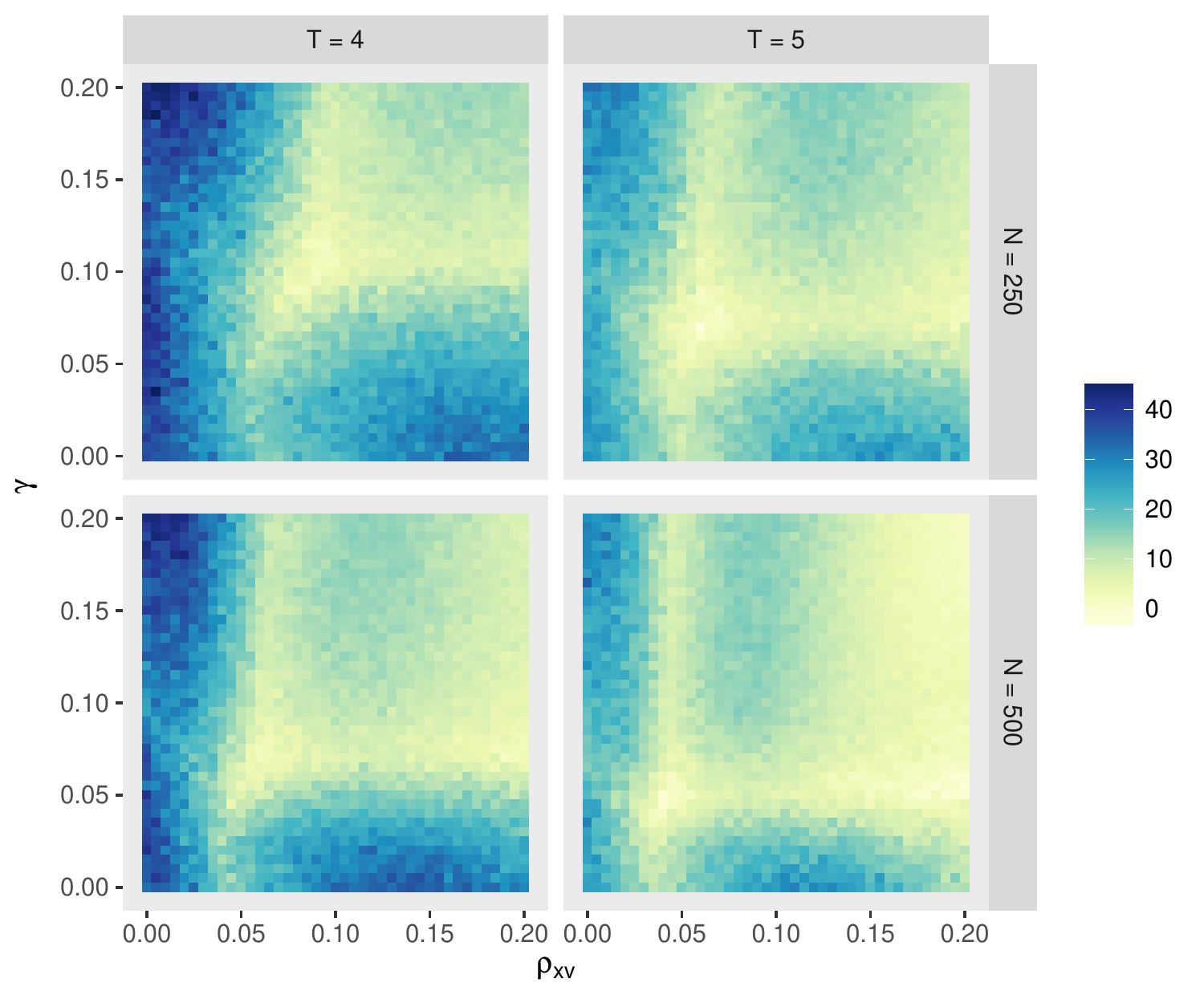}
\caption{RMSE of GFIC relative to Oracle Estimator in the Simulation from Section \ref{sec:Dpanel_sim_SR}}
\label{fig:RMSE_rel_oracle}
\end{figure}

In this section we present figures to complement Table \ref{tab:Dpanel_RMSE} from Section \ref{sec:Dpanel_sim_SR}.
Figure \ref{fig:best} colors each region of the parameter space according to which of the estimators of $\theta$ -- $\text{LP}$, $\text{LS}$, $\text{P}$ or $\text{S}$ -- yields the lowest finite-sample RMSE.
The saturation of a color indicates the relative difference in RMSE of the lowest RMSE estimator at that point measured against the \emph{second lowest} RMSE estimator.
Darker indicates a larger advantage for the first-best estimator while lighter values indicate a smaller advantage. 
Figure \ref{fig:GFIC_rel_LP} depicts the relative different between the RMSE of the GFIC and that of the true specification, LP, expressed in percentage points.
Red indicates that the GFIC has the lower RMSE, blue that LP has the lower RMSE, and white that the RMSE values are the same. 
Darker colors indicate a larger difference.
Figure \ref{fig:RMSE_rel_oracle} compares the RMSE of the GFIC to that of the oracle procedure that uses whichever fixed specification -- LP, LS, P, or S -- yields the lowest finite sample MSE at a give point in the parameter space.
As in Figure \ref{fig:GFIC_rel_LP}, the comparison is one of relative RMSE in percentage points.
But, as the GFIC can by definition can never have a lower finite-sample MSE than the oracle estimator, the color scale used in this figure is different.
The remaining figures in this section compare the RMSE of the GFIC to that of the other selection procedures: GMM-AIC, GMM-BIC, etc.

\begin{figure}
\centering
\includegraphics[scale = 0.8]{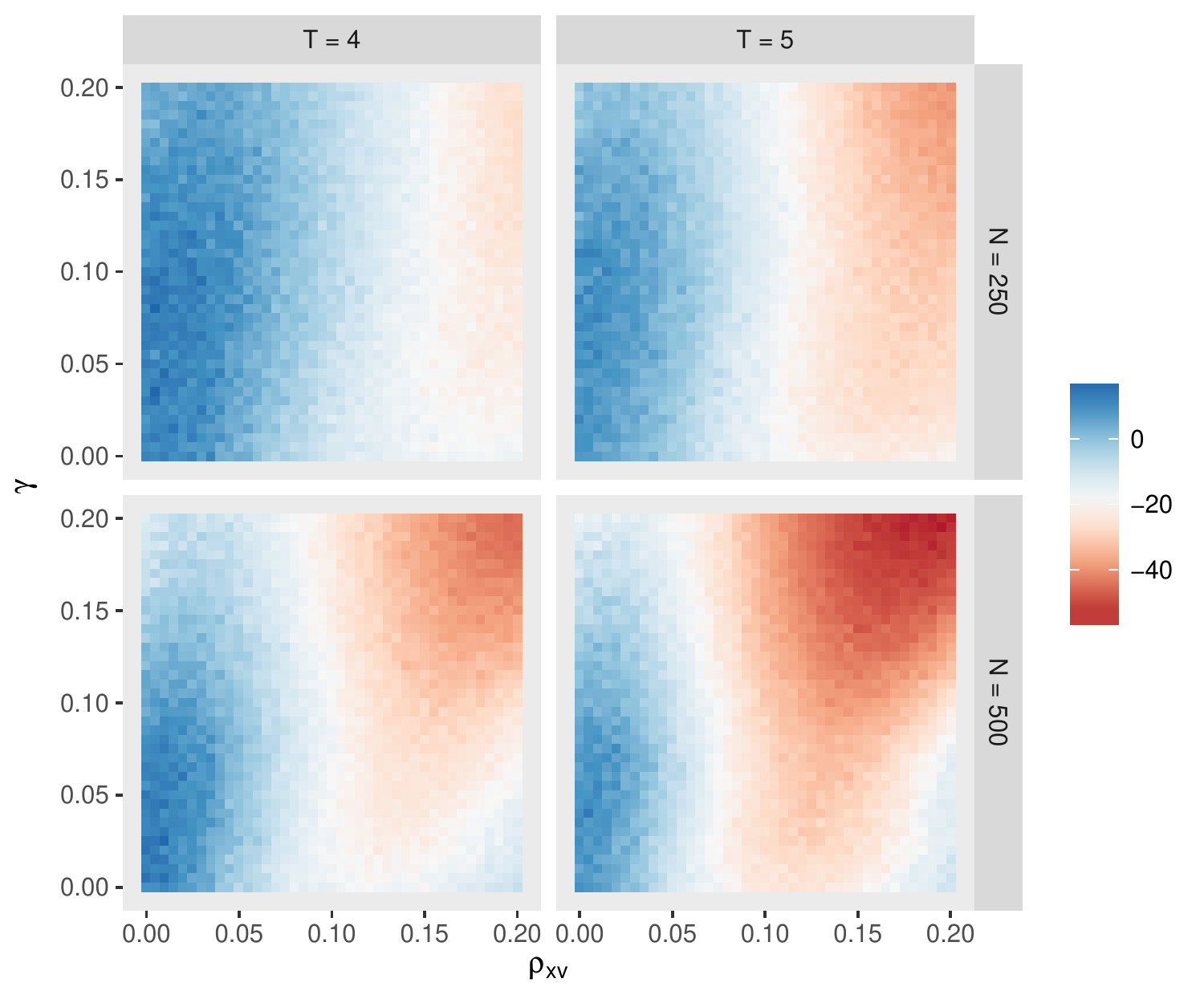}
\caption{RMSE of GFIC relative to GMM-AIC in the Simulation from Section \ref{sec:Dpanel_sim_SR}}
\end{figure}
\begin{figure}
\centering
\includegraphics[scale = 0.8]{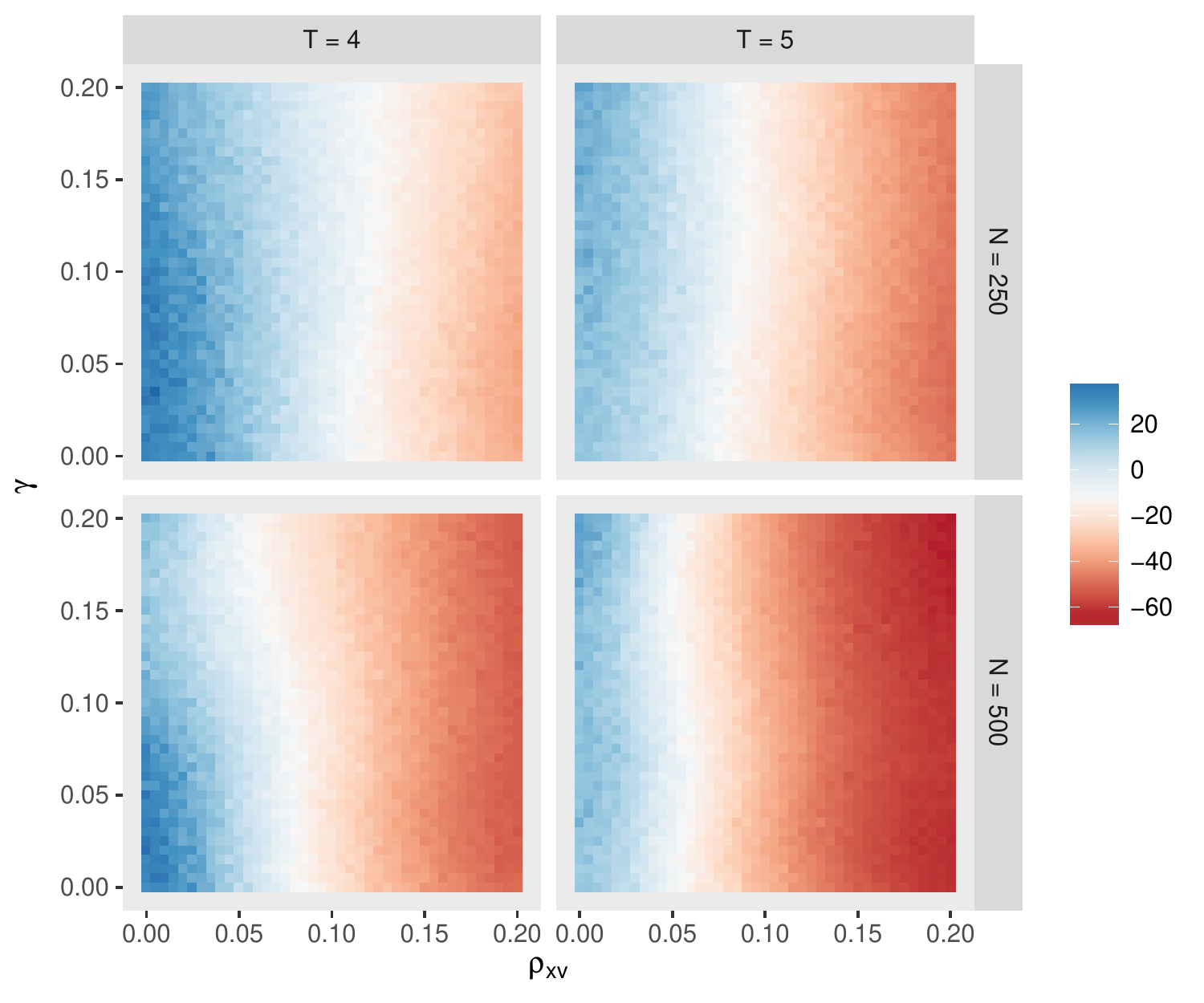}
\caption{RMSE of GFIC relative to GMM-BIC in the Simulation from Section \ref{sec:Dpanel_sim_SR}}
\end{figure}
\begin{figure}
\centering
\includegraphics[scale = 0.8]{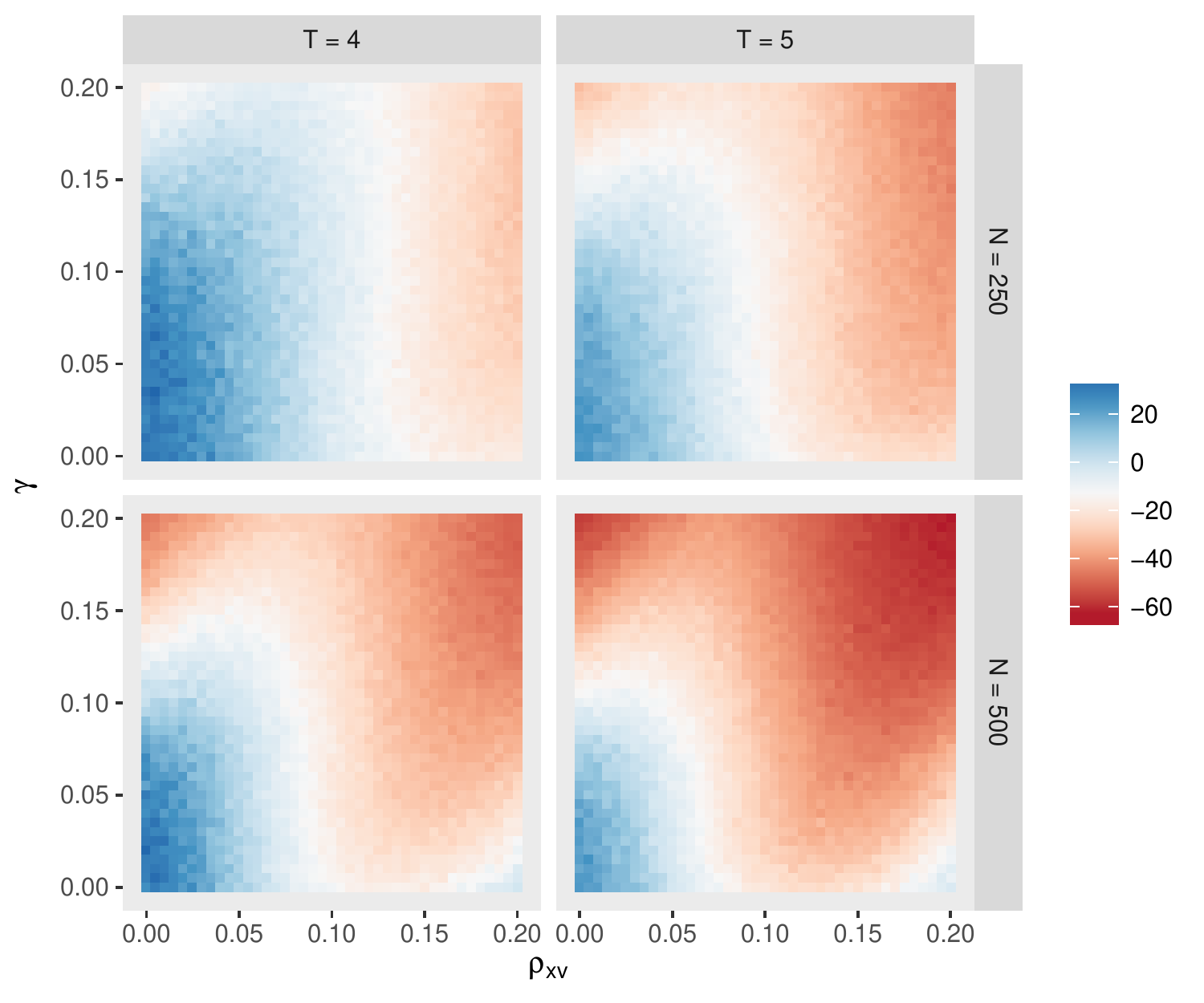}
\caption{RMSE of GFIC relative to 5\% Downward J-test in the Simulation from Section \ref{sec:Dpanel_sim_SR}}
\end{figure}
\begin{figure}
\centering
\includegraphics[scale = 0.8]{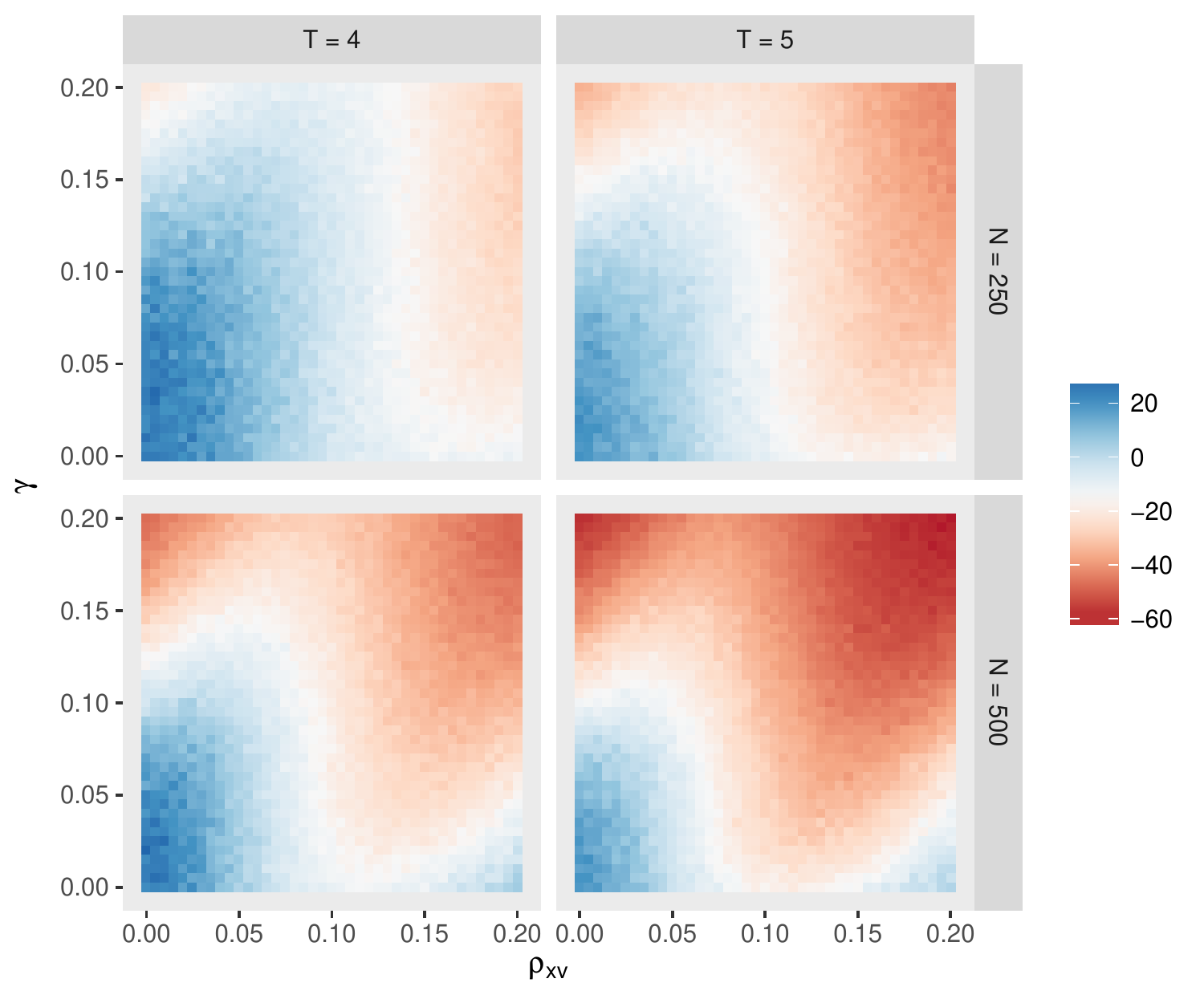}
\caption{RMSE of GFIC relative to 10\% Downward J-test in the Simulation from Section \ref{sec:Dpanel_sim_SR}}
\end{figure}

\begin{figure}
\centering
\includegraphics[scale = 0.8]{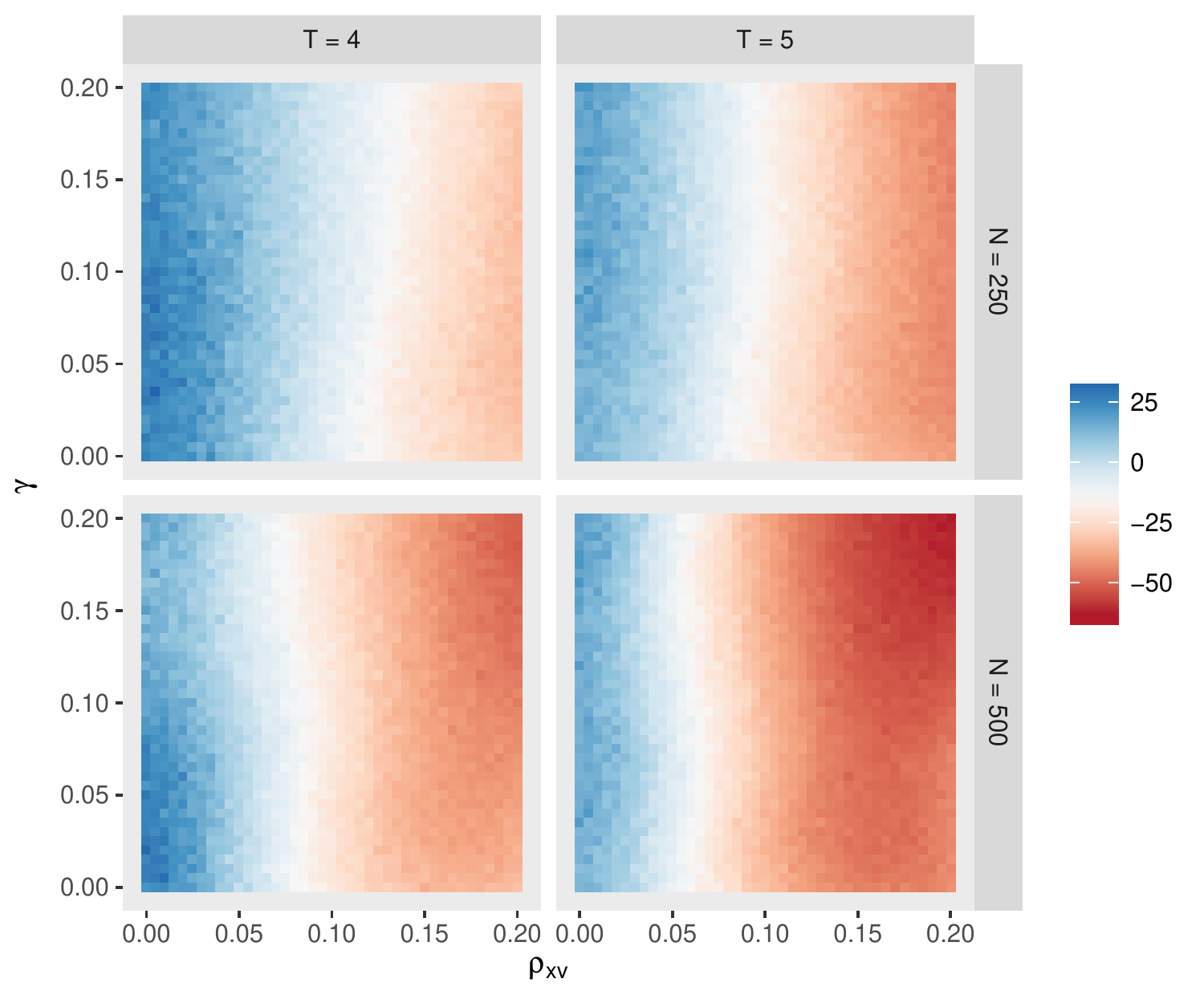}
\caption{RMSE of GFIC relative to GMM-HQ in the Simulation from Section \ref{sec:Dpanel_sim_SR}}
\end{figure}


\newpage
\section{Supplementary Results for the Empirical Example}
\label{sec:append_empirical}

Table \ref{tab:cigarettesappend} presents additional empirical results to supplement those discussed in Section \ref{sec:cigarettes}: \ref{tab:cigarettesLongerappend} is based on data for the 1963-1974 sub-sample ($T = 12$) while \ref{tab:cigarettesFullappend} is based on data for the full 1963--1992 sample ($T=30$).
The results in \ref{tab:cigarettesShortappend} and \ref{tab:cigarettesLongappend} are the same as those in Table \ref{tab:cigarettes} and are included here for ease of comparison.
Although there is some variation  in the magnitudes of coefficient estimates across sub-samples, the basic pattern of results is similar.
In each of the longer samples ($T=11, 12$ or $30$), the GFIC ranks specification LP as the best and specification P as the second best.
With enough time periods available for estimation, the reduction in  variance from using specifications LS, P, and S becomes negligible and is hence outweighed by any bias that they may induce.

\begin{table}[h]
  \centering
    \begin{subtable}[h]{0.45\textwidth}
        \centering
     \caption{1975--1980 ($T=6$)}
     \label{tab:cigarettesShortappend}
     \begin{tabular}{lrrrr}\hline\hline 
         & \multicolumn{1}{c}{$\text{LP}$} & \multicolumn{1}{c}{$\text{LS}$} 
          & \multicolumn{1}{c}{$\text{P}$} & \multicolumn{1}{c}{$\text{S}$}\\
        \hline
        $\widehat{\theta}$ & -0.68 & -0.32 &  -0.28 &  -0.37\\
        Var.\ & 0.16 &  0.02 & 0.07 & 0.01\\ 
        Bias$^2$ & \multicolumn{1}{c}{---} & -4.20 & 0.01 & -3.56 \\
        GFIC  & 0.16 & -4.18 & 0.08  & -3.54  \\
        GFIC+  & 0.16 &  0.02 & 0.08  & 0.01    \\
        \hline
      \end{tabular}
    \end{subtable}
    ~
    \begin{subtable}[h]{0.45\textwidth}
      \centering
     \caption{1975--1985 ($T=11$)}
     \label{tab:cigarettesLongappend}
     \begin{tabular}{lrrrr}\hline\hline 
         & \multicolumn{1}{c}{$\text{LP}$} & \multicolumn{1}{c}{$\text{LS}$} 
          & \multicolumn{1}{c}{$\text{P}$} & \multicolumn{1}{c}{$\text{S}$}\\
        \hline
        $\widehat{\theta}$ & -0.30 & -0.26 &  -0.38 &  -0.28\\
        Var.\ & 0.06 & 0.01 & 0.05 & 0.01\\ 
        Bias$^2$ & \multicolumn{1}{c}{---} & 2.21 & 0.03 & 1.29\\
        GFIC  &0.06 & 2.22 & 0.08 &  1.30\\
        GFIC+  & 0.06 & 2.22 & 0.08  & 1.30  \\
           \hline
      \end{tabular}     
    \end{subtable}
     \begin{subtable}[h]{0.45\textwidth}
        \centering
     \caption{1963--1974 ($T=12$)}
     \label{tab:cigarettesLongerappend}
     \begin{tabular}{lrrrr}\hline\hline 
         & \multicolumn{1}{c}{$\text{LP}$} & \multicolumn{1}{c}{$\text{LS}$} 
          & \multicolumn{1}{c}{$\text{P}$} & \multicolumn{1}{c}{$\text{S}$}\\
        \hline
        $\widehat{\theta}$ & -0.31 & -0.52 &  -0.16 &  -0.51\\
        Var.\ & 0.03 &  0.00 & 0.04 & 0.00\\ 
        Bias$^2$ & \multicolumn{1}{c}{---} & 1.15 & 0.61 & 1.28 \\
        GFIC  & 0.03 & 1.15 & 0.66  & 1.28   \\
        GFIC+  & 0.03 &  1.15 & 0.66  & 1.28    \\
        \hline
      \end{tabular}
    \end{subtable}
     ~
    \begin{subtable}[h]{0.45\textwidth}
      \centering
     \caption{1963--1992 ($T=30$)}
     \label{tab:cigarettesFullappend}
     \begin{tabular}{lrrrr}\hline\hline 
         & \multicolumn{1}{c}{$\text{LP}$} & \multicolumn{1}{c}{$\text{LS}$} 
          & \multicolumn{1}{c}{$\text{P}$} & \multicolumn{1}{c}{$\text{S}$}\\
        \hline
        $\widehat{\theta}$ & -0.15 & -0.38 &  -0.07 &  -0.38\\
        Var.\ & 0.01 & 0.00 & 0.01 & 0.00\\ 
        Bias$^2$ & \multicolumn{1}{c}{---} & 2.36 & 0.35 & 2.18\\
        GFIC  & 0.01 & 2.36 & 0.36 &  2.18\\
        GFIC+  & 0.01 & 2.36 & 0.36  & 2.18  \\
           \hline
      \end{tabular}     
    \end{subtable}
    \caption{Estimates and GFIC values for the price elasticity of demand for cigarettes example from Section \ref{sec:cigarettes} under four alternative specifications. GFIC+ gives an alternative version of the GFIC in which a negative squared bias estimate is set equal to zero.}
    \label{tab:cigarettesappend}
\end{table}

\end{document}